\documentclass[twocolumn]{aastex62}
\usepackage{CJK}
\usepackage{amsbsy}
\usepackage{multirow}
\usepackage{bm}
\usepackage{ulem}

\def\kms{{\,\rm km\,s^{-1}}}

\def\msunh{\,{\rm h^{-1} M}_\odot}
\def\mpc{\,{\rm Mpc}}
\def\mpch{\,{\rm h^{-1}  Mpc}}

\def\nsate{\,{\rm N_{sate}}}

\defcitealias{2020ApJ...898L..15B}{BS2020}

\submitjournal{ApJ}

\shorttitle{lopsided satellite distribution}
\shortauthors{Peng Wang et al.}

\begin{document}
\begin{CJK*}{UTF8}{gbsn}
\title{The lopsided distribution of satellites of isolated central galaxies}
\correspondingauthor{Peng Wang}
\email{pwang@aip.de}

\author[0000-0003-2504-3835]{Peng Wang (王鹏)}
\affil{Leibniz-Institut f\"ur Astrophysik Potsdam, An der Sternwarte 16, D-14482 Potsdam, Germany}

\author{Noam I. Libeskind (李诺恩)}
\affil{Leibniz-Institut f\"ur Astrophysik Potsdam, An der Sternwarte 16, D-14482 Potsdam, Germany}
\affil{University of Lyon; UCB Lyon 1/CNRS/IN2P3; IPN Lyon (IPNL), France}

\author[0000-0002- 9197-9300]{Marcel S. Pawlowski}
\affil{Leibniz-Institut f\"ur Astrophysik Potsdam, An der Sternwarte 16, D-14482 Potsdam, Germany}

\author[0000-0002-5458-4254]{Xi Kang (康熙)}
\affil{Zhejiang University-Purple Mountain Observatory Joint Research Center for Astronomy, Zhejiang University, Hangzhou 310027, China}
\affil{Purple Mountain Observatory, No. 10 Yuan Hua Road, 210034 Nanjing, China}

\author{Wei Wang (王伟)}
\affil{Purple Mountain Observatory, No. 10 Yuan Hua Road, 210034 Nanjing, China}
\affil{School of Astronomy and Space Science, University of Science and Technology of China, Hefei 230026, Anhui, China}

\author{Quan Guo (郭铨)}
\affil{Shanghai Astronomical Observatory, Nandan Road 80, Shanghai 200030, China}

\author[0000-0002-5249-7018]{Elmo Tempel}
\affil{Tartu Observatory, University of Tartu, Observatooriumi 1, 61602 T\~oravere, Estonia}

\begin{abstract}
Satellites are not randomly distributed around their central galaxies but show polar and planar structures. In this paper, we investigate the axis-asymmetry or lopsidedness of satellite galaxy distributions around isolated galaxies in a hydrodynamic cosmological simulation. We find a statistically significant lopsided signal by studying the angular distribution of the satellite galaxies' projected positions around isolated central galaxies in a  two-dimensional plane. The signal is dependent on galaxy mass, color and large-scale environment. Satellites that inhabit low-mass blue hosts, or located further from the hosts show the most lopsided signal. Galaxy systems with massive neighbors exhibit stronger lopsidedness. This satellite axis-asymmetry signal also decreases as the universe evolves. Our findings are in agreement with recent observational results, and they provide useful perspective for studying galaxy evolution, especially on the  satellite accretion, internal evolution and interaction with the cosmic large-scale structure.

\end{abstract}

\keywords{Cold dark matter -- Dwarf galaxies -- Large-scale structure of the universe -- Astrostatistics}

\section{Introduction}\label{sec:intro}

It is perhaps a pause for thought that the most abundant objects in the universe are its smallest entities: dwarf galaxies. According to the widely favored paradigm of structure formation (known as the $\Lambda$CDM model), small objects are the first to collapse in early Universe. Larger objects, like Milky Way-type galaxies, are built via the merging of these dwarfs. The merging process is not instantaneous, and small dwarfs become satellite galaxies, orbiting around larger bodies until  they merge with the central object or being disrupted by dynamical processes, such as tidal heating and disruption.

So long as satellite galaxies exist, their phase space distributions can be used to trace the gravitational potential of the host halo they inhabit. For example, the satellite distribution can be used to trace the shape of the host halo and can thus be used to estimate a ``dynamical'' mass of the halo \citep{2004MNRAS.352.1302V}. The ideal application of such techniques is more suitable for spherical halos that can be described by a one-dimensional (radial) density profile, e.g. NFW. However, it is well known that nature does not obey such simplifying assumptions.

Much work has been done to show the anisotropic (albeit axis-symmetric) distribution of satellite galaxies, first noted half a century ago \citep[][and reference within]{1968PASP...80..252S, 1969ArA.....5..305H}.  After decades of debate \citep[e.g.,][]{2004MNRAS.348.1236S, 2005ApJ...628L.101B, 2005MNRAS.363..146L, 2006MNRAS.369.1293Y, 2007MNRAS.378.1531K, 2007MNRAS.376L..43A, 2008MNRAS.390.1133B, 2010ApJ...709.1321A, 2018ApJ...859..115W, 2019A&A...628A..31G, 2019RAA....19..181Z, 2020ApJ...893...87T} the preferential distribution of satellite galaxies around the major axis of their host central galaxy has eventually been widely accepted. The alignment signal is dominated by red satellites of red central galaxies, while blue satellites of blue central galaxies have almost random distributions. What's more, the anisotropic and kinematically correlated distribution of satellites, the so-called ``plane of satellites''  issue, is still under debate \citep{2017ApJ...843...62M}. Such co-rotating, spatially thin distributions of satellites have been observed the Milky Way \citep[hereafter MW][]{2012MNRAS.423.1109P,  2015MNRAS.452.3838C, 2019MNRAS.488.1166S}, Andromeda \citep[hereafter M31][]{2013Natur.493...62I} and Centaurus A  \citep[][]{2015ApJ...802L..25T,  2018Sci...359..534M} in the local universe.

The aspherical distributions mentioned above remain axis-symmetric or at least are categorized as such. However, just like the ellipsoidal density profile of \cite{2002ApJ...574..538J} is a better approximation of nature than a spherical profile such as NFW, the approximation of axis-symmetry could also be improved. In fact, when examining the satellite galaxies in the Local Group,  it is noted that 7 out of the 11 brightest satellites of the Milky Way are located in one hemisphere. In M31, the situation is more drastic with 21 out of 27 \citep{2013ApJ...766..120C} \citep[or 23 out of 30 stated by][]{2020MNRAS.492..456W} satellites on the same side of the host. Since the MW and M31 are a galaxy pair, it is found that there are 25 satellites which lie in the region between the two galaxies.

Inspired by the  ``lopsided'' satellites distribution (hereafter LSD) in the Local Group, \cite{2016ApJ...830..121L} investigated this phenomenon in galaxy pairs with similar magnitudes and separations as the Local Group. Using SDSS DR10 \citep{2000AJ....120.1579Y,  2014ApJS..211...17A}, they found that about 8\% more satellites than expected from a uniform distribution reside in a region defined by a $20^{\circ}$ opening angle between galaxy pairs. Following up on \cite{2016ApJ...830..121L}, \cite{2017ApJ...850..132P} examined whether the observed lopsided signal is consistent with prediction from $\Lambda$CDM cosmological N-body simulations and they found that such setups are indeed expected. Recently, \cite{2019MNRAS.488.3100G} examined the origin of these LSDs in simulations, attributing their existence to the accretion of satellites along filaments that connect the two members of a galaxy pair. The lopsided signal is driven by satellites on their first approaching orbits, thus lopsided systems represent dynamically active environments. As satellites are accreted and become virialized, their distribution moves away from lopsided and towards anisotropic albeit axis-symmetric distributions.

Studies that focusing on isolated (i.e., unpaired) host galaxies have also (often serendipitously) found LSDs. For example, 7 out of 8 confirmed bright satellites of M101 reside on one side. M101 displays a LSD even if all possible satellite candidates (including unconfirmed satellites) are considered \citep{2019ApJ...885..153B, 2020ApJ...893L...9B}. Beyond this ``anecdotal'' case, at least one observational study \citep[][hereafter BS2020]{2020ApJ...898L..15B} examined this phenomenon on a statistical scale by using the NASA-Sloan Atlas (NSA) catalog. \citetalias{2020ApJ...898L..15B} convincingly demonstrated that LSDs are not limited to paired galaxies but also in isolated galaxies with a color dependence:  bluer hosts have more pronounced lopsided satellite systems, consistent with the picture painted in \cite{2019MNRAS.488.3100G}, who suggested that dynamically younger systems have more pronounced LSDs.

To test whether the observed lopsided signal for isolated galaxies can be confirmed in the standard ΛCDM cosmological model, we investigate this phenomenon in the start-of-the-art hydrodynamic cosmological simulation Illustris-TNG. Our paper is organized as follows. Section \ref{sec:method} presents the simulation data and the method to quantify the degree of LSD and its statistical significance. In Section \ref{sec:result}, we show the results, including the effects of various properties of the hosts, the satellites, and their surrounding large-scale cosmic web. Our conclusions and discussion are presented in Section \ref{sec:sum_dis}.

\section{Data and Methodology}
\label{sec:method}
We use the publicly available start-of-the-art hydrodynamic cosmological simulation Illustris-TNG. The adopted cosmological parameters employed by Illustris-TNG are taken from Planck \citep{2016A&A...594A..13P}, namely: $\Omega_\Lambda=0.6911$ $\Omega_{\rm m}=0.3089$, ${\rm H_0}= 100 h \rm{\kms Mpc^{-1}}$ with $h=0.6774$. The set of TNG simulations contains three box size runs: 50, 100, and 300 $\rm Mpc^{3}$. In this work, we employ the hydrodynamic run with the largest box size, the TNG300-1 \citep{ 2018MNRAS.480.5113M}, to ensure enough central galaxies. The mass resolution achieved is $m_{\rm p}\sim 5.9\times10^{7}M_{\odot}$ per particle. Dark matter halos are identified by the standard Friend-of-Friend algorithm \citep{1985ApJ...292..371D}. The SUBFIND algorithm \citep{2001MNRAS.328..726S, 2009MNRAS.399..497D} is applied to each FoF group to identify gravitationally bound structures. For each halo, the galaxy inhabiting the most massive SUBFIND (sub)halo is regarded as the central galaxy. All other substructures are considered satellites. Galaxy positions are assigned to the position of the most bound particle of the corresponding subhalo.

\begin{figure*}[!ht]
\plotthree{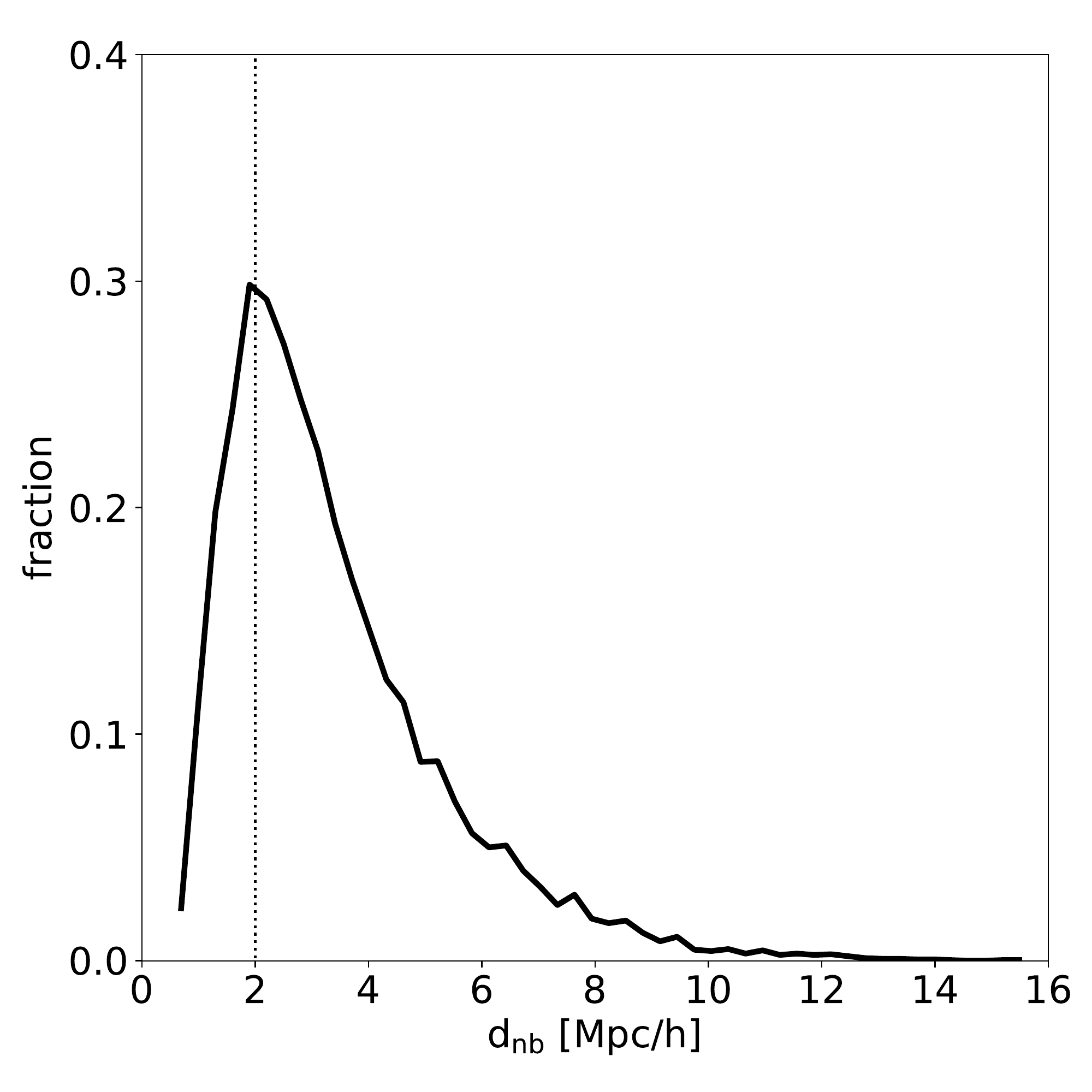}{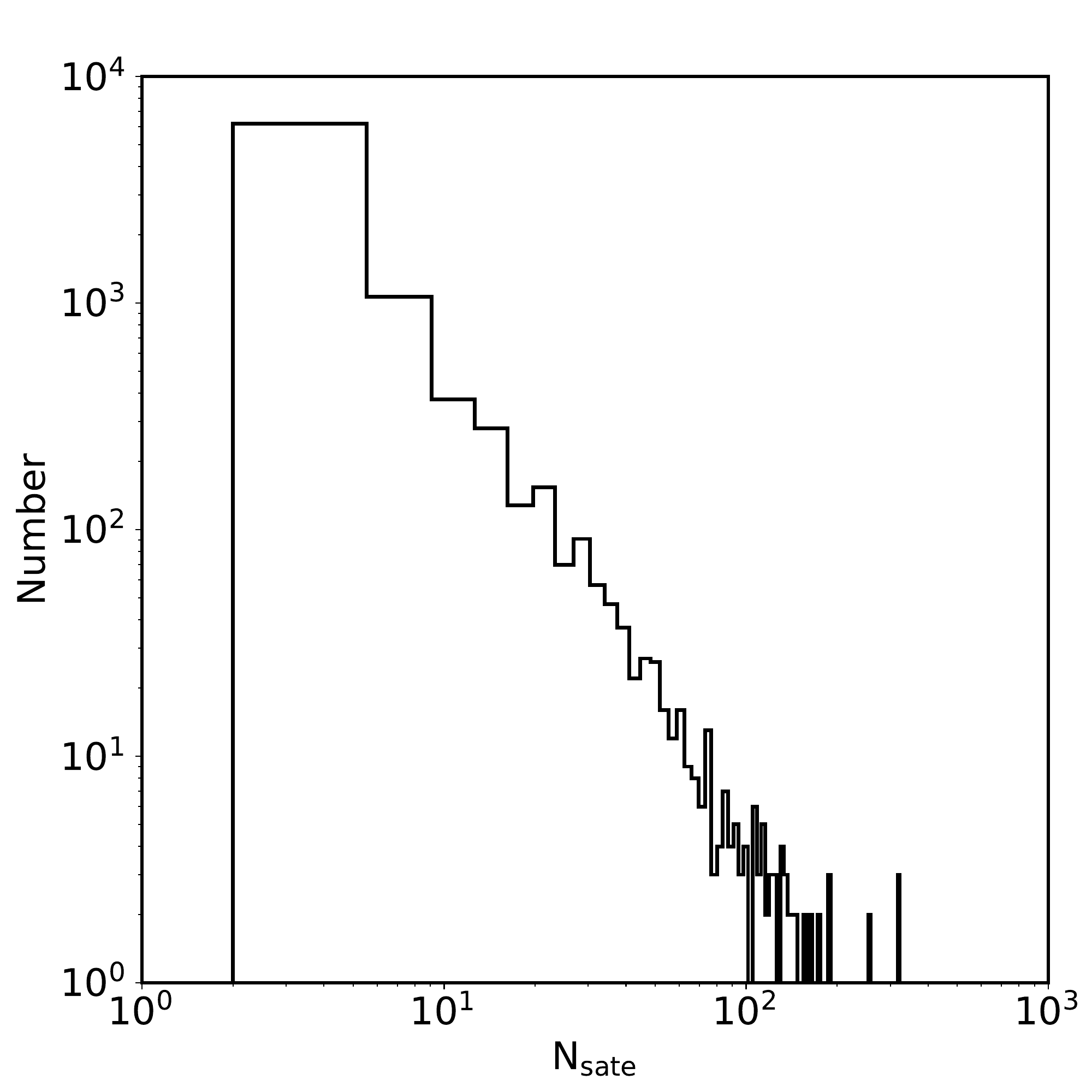}{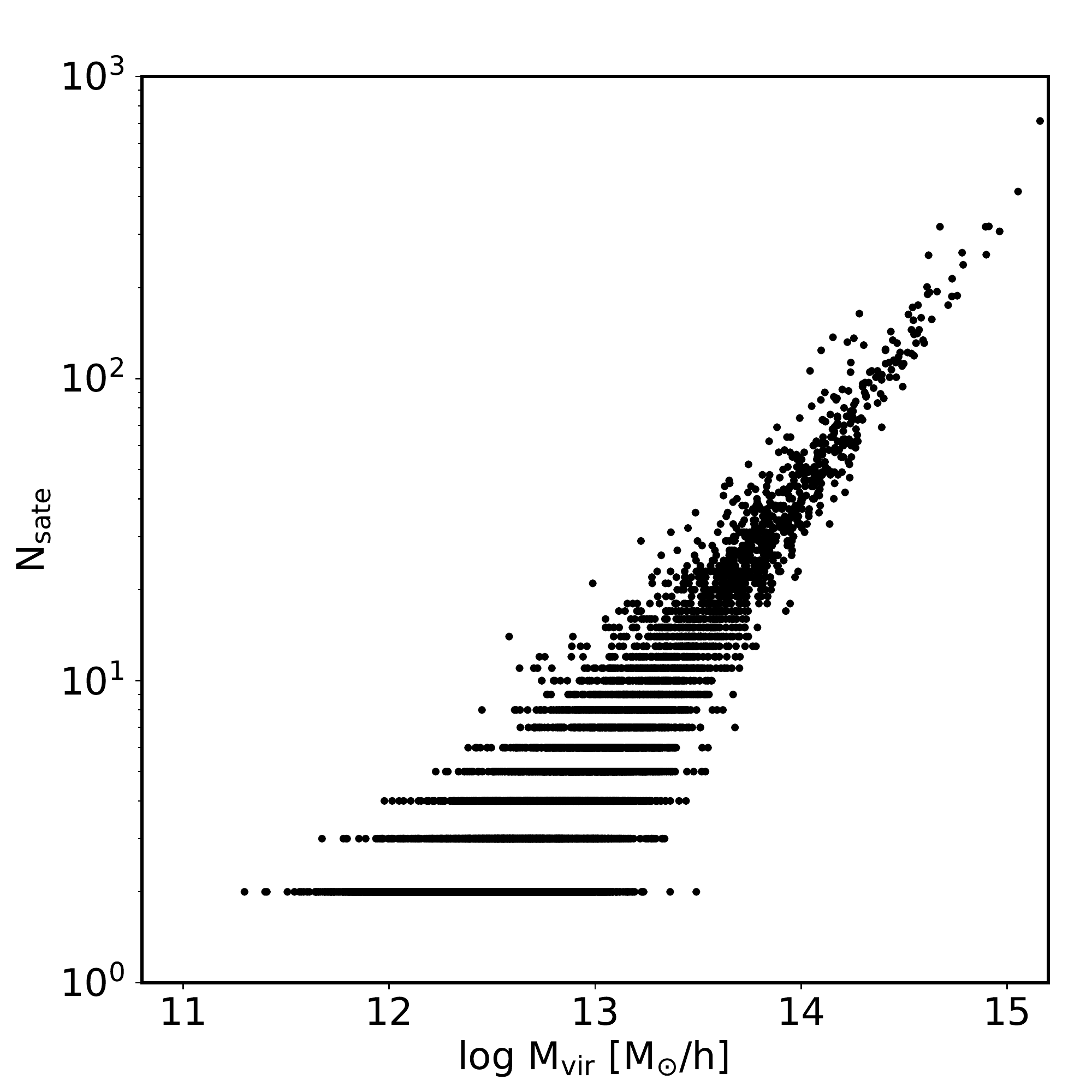}
\caption{{\bf Selection criteria for isolated central galaxies and their properties ($\nsate$ and $\rm M_{vir}$).} 
{\bf Left panel}: the distribution of the distance between a given central to the closest neighbor central. 
{\bf Middle panel}: the distribution of $\nsate$ of selected isolated central galaxies. 
{\bf Right panel}: the relation between halo mass $\rm M_{vir}$ and  $\nsate$. }
\label{fig:sample}
\end{figure*}

\begin{figure}[!ht]
\plotbigone{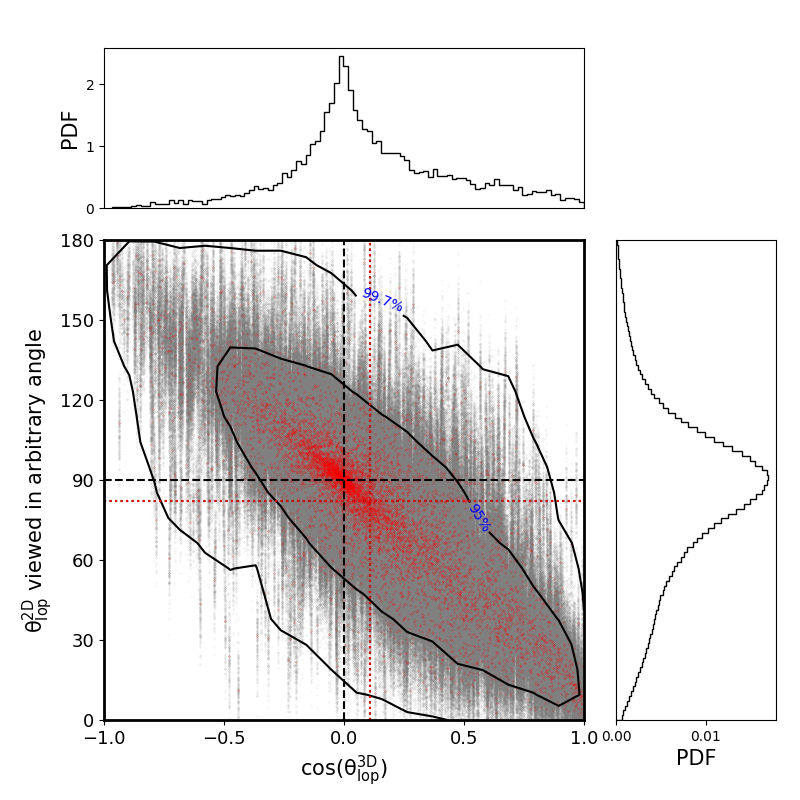}
\caption{{\bf The correlation between the lopsided angles measured in 3D and 2D of each system.} Red dots represent the 8,724 systems we selected from the simulation (projected in the X-Y plane), and the gray dots represent 1,000 random projections (the distribution of $\cos\theta^{3D}_{lop}$ and $\theta^{2D}_{lop}$ is shown in the top and right panel, respectively) of the simulation sample. The black dotted lines represent the points of $\cos(\theta^{3D}_{lop})=0.0$ and $\theta^{2D}_{lop}$=90$^\circ$. Red and black (at same values) lines represent the mean of the simulation sample and random samples, respectively. The black contour lines represent 95\% and 99.7\% of the number fraction of random trials.}
\label{fig:coslop_2d_3d}
\end{figure}

\begin{figure*}[!ht]
\plottwo{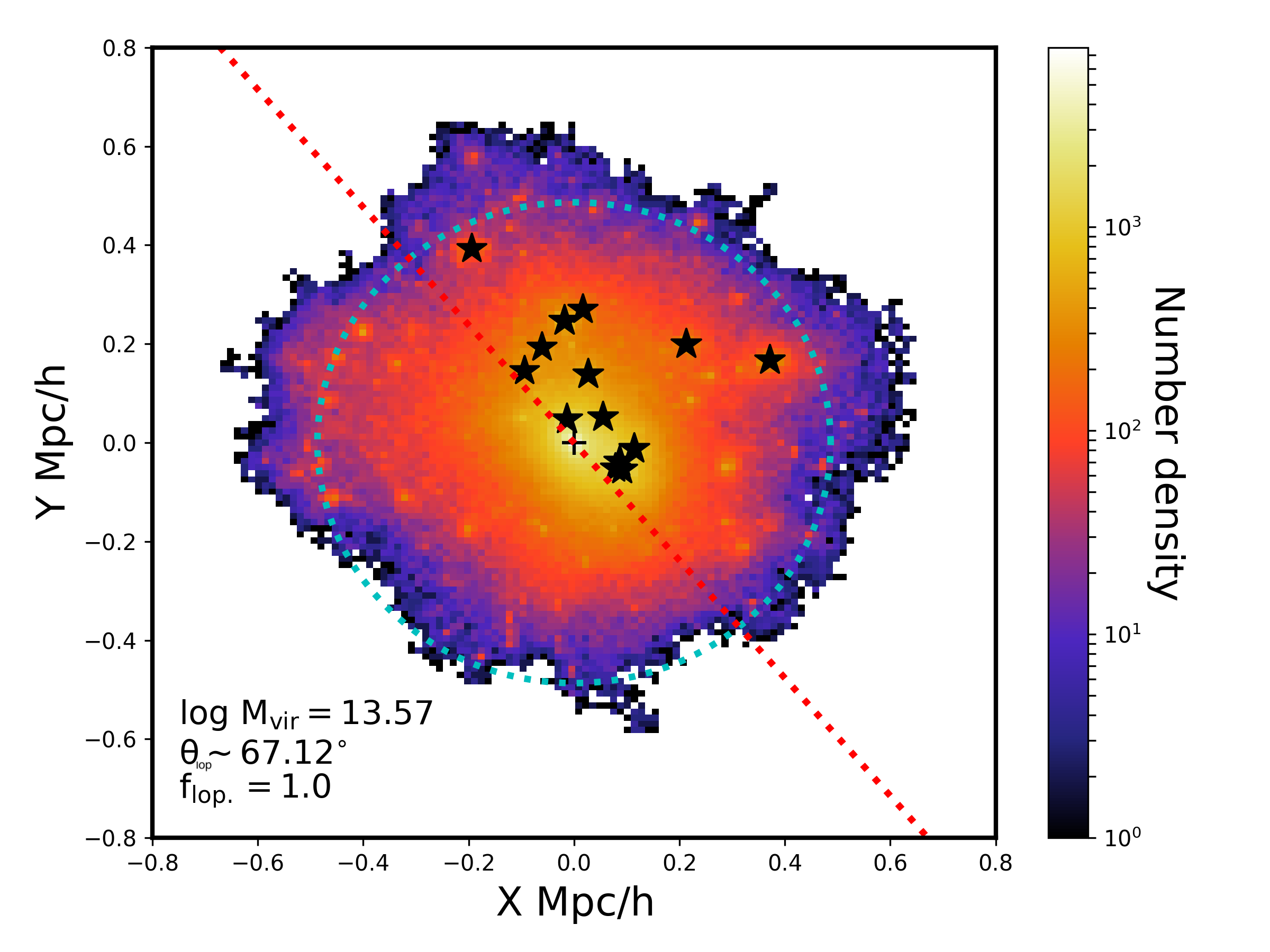}{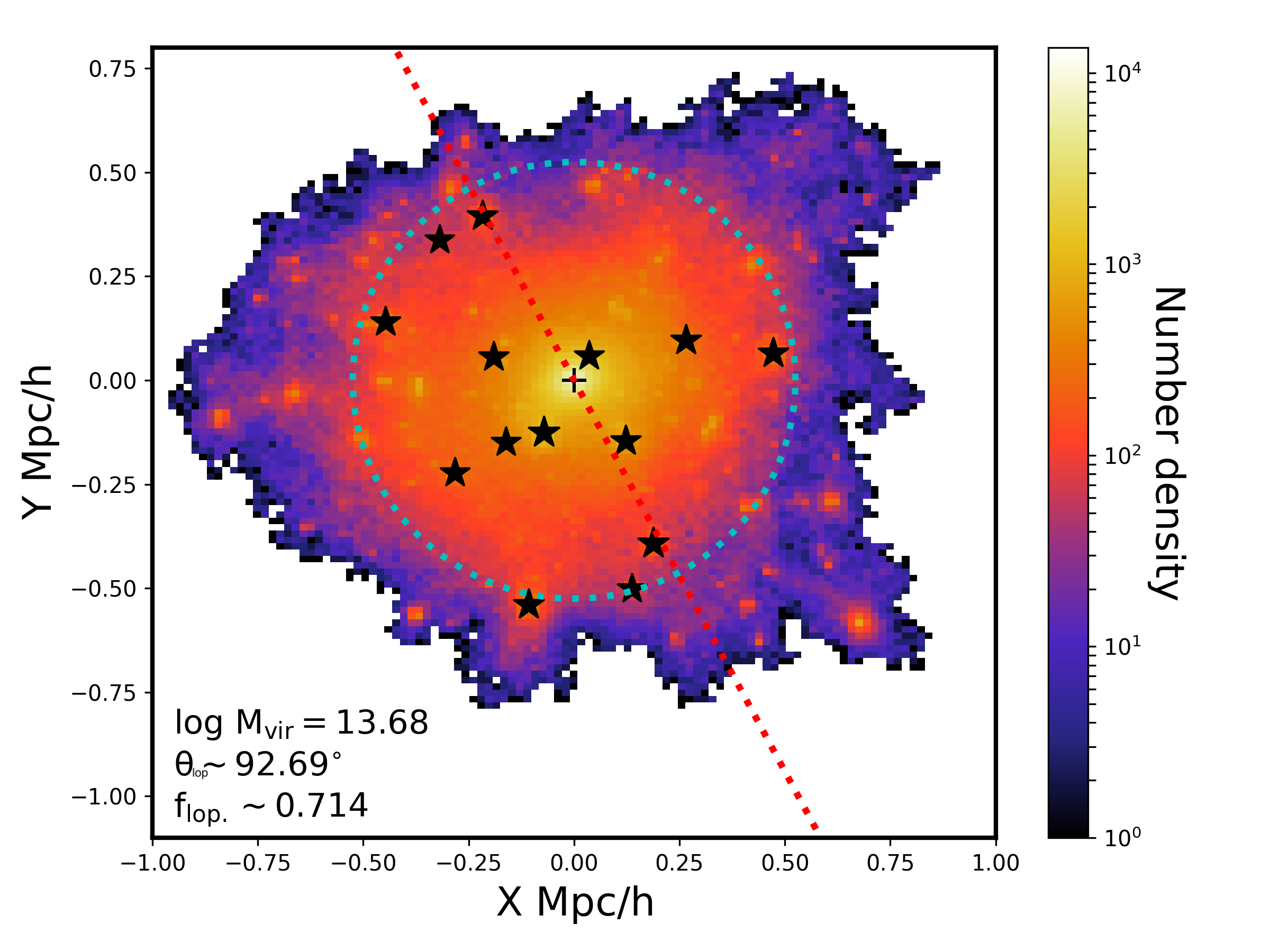}
\caption{{\bf Illustration of two cases in similar halo mass and satellite number but with different lopsidedness.} 
A case of satellite galaxies completely distributed in one side (lopsided angle $\rm \theta_{lop}$=$67.12^{\circ}$ and lopsidedness fraction $\rm f_{lop.}$=$1.0$) is displayed in the {\bf left panel}, and the other case of nearly isotropic distribution (lopsided angle $\rm \theta_{lop}$=$92.69^{\circ}$ and lopsidedness fraction $\rm f_{lop.}$=$0.714$) is shown in the {\bf right panel}. 
The background shows the density field of dark matter particles of the halo and halo virial radius is shown in the dashed cyan circle. Black star symbols indicate the satellite positions.  The split corresponding to $\rm f_{lop.}$ is shown with a red dotted line. }
\label{fig:lopsided_case}
\end{figure*}

Given the simulation resolution, we choose central and satellite galaxies with stellar mass of more than $10^{11}$ and $10^{9}$, in units of $\msunh$, respectively. 
Note that our sample of central galaxies is more massive (brighter) than those in the observational study \citepalias{2020ApJ...898L..15B}.
We use a simple and effective definition for isolated central galaxies: A galaxy is considered ``isolated'' if the closest massive halo (whose stellar mass is $> 10^{11}\msunh$) is further than $2 \mpch$ away. Note that for systems close to the edge of the simulation box, the periodic boundary condition is considered.
This isolation criterion is mainly motivated by the following consideration: the distribution of the distance between central galaxies (with mass $> 10^{11}\msunh$) peaks at $2 \mpch$ (see left panel of Fig.~\ref{fig:sample}). In addition, \cite{2016ApJ...830..121L} found the lopsided signal for galaxy pairs separated by more than $1.5 \mpc$ is driven by stochasticity. 
Our final sample consists of 8,724 isolated centrals with each host containing at least two satellites. The mean (median) number of satellites per host is 8 (3), while the total number of satellites in the sample is 69,694. 
We present in Fig.~\ref{fig:sample} (middle panel) the distribution of the number of satellites, $\nsate$ per host. The right panel of Fig.~\ref{fig:sample} shows the correlation between halo viral mass, $\rm M_{vir}$, and $\nsate$. Clearly, more massive halos contain more satellites.

The ``lopsidedness'' of a system of satellites can be quantified by examining the angle $\theta^{ij}$ (in which $i,j=1,2,...,\nsate$ but $i<j$) between each pair of satellites ($i$ \& $j$) in a given host. An isotropic or axis-symmetric distribution of satellites has a uniform distributions of $\theta^{ij}$. Departures from a uniform distribution that are statistically significant are thus a measure of the axis-asymmetry of these satellite distributions. In this work, we use two measures of the lopsidedness of a system: the first is the above mentioned $\theta^{ij}$ \citep[same as used in][]{2020ApJ...898L..15B} and the second is the lopsided angle $\theta_{\rm lop}$=$\langle\theta^{ij}\rangle$, which is the mean of the angles $\theta^{ij}$ for one given system. 
Statistically, in the case of isotropic satellite distributions, the expected mean value of the $\theta_{\rm lop}$ in 2D is 90 degrees (in 3D $\cos \theta_{\rm lop}$=0). If $\theta_{\rm lop}< 90^{\circ}$ ($\cos\theta_{\rm lop}>0$),  we refer to a lopsided distribution, which means more satellites are located on one side.  
However, we note that not all lopsided distributed systems have to have $\theta_{\rm lop}< 90^{\circ}$. Some systems with small $\nsate$ have  $\theta_{\rm lop}> 90^{\circ}$ (see  App~\ref{sec:flop_angle}).
Thus, $\theta_{\rm lop}$ is not a perfect indicator to measure the lopsidedness in the satellite distribution for an individual system with a small $\nsate$.

The observational study \citepalias{2020ApJ...898L..15B} examined the projected angular position of satellites with respect to their hosts. Unlike the observed universe, satellite systems can be viewed from any arbitrary angle in the simulation. One may ask whether the viewing angle affects the LSD signal. To address this question, we calculate the lopsided angle both in 3D and 2D space. For the 2D case, we viewed each satellite system at 1000 random angles.  As shown in Fig.~\ref{fig:coslop_2d_3d}, we examined the correlation of the lopsided angle between 3D and 2D space. The red points display the 8,724 systems, and grey dots show random trials with 1000 random viewing angles per system. We found that different viewing angles do affect the lopsided signal, but statistically, the 2D and 3D signals have a very strong linear relationship. Therefore, it is possible to use any projection angle to study the LSD. In the following, we show the results calculated in one arbitrarily chosen  projection of the simulation: the X-Y plane.

The significance of a given lopsided signal is sensitive to the number of satellite galaxies, $\nsate$ per system. Therefore a statistical check is needed to examine whether the measured signal is inconsistent with stochastic processes and, if so, at what level. A simple Monte-Carlo test assesses the statistical significance. For each system, the angular distribution of the $\nsate$ satellites is randomized. We perform this randomization test 10,000 times per system, and the lopsided angle $\rm \theta_{\rm lop}^{R}(=\langle \theta^{ij}_R\rangle)$ is determined for each randomized satellite distribution.  With these, a mean lopsided angle $\langle \rm \theta_{lop}^{R} \rangle$ and the standard deviation $\rm \sigma(\theta_{lop}^{R})$  of the randomized trials can be estimated for each system. The statistical significance of any measured lopsided angle $\theta$ can then be determined by 
\begin{equation}
significance = \rm \frac{|\theta_{\rm lop}-\langle \theta_{\rm lop}^{R} \rangle|}{\sigma(\theta_{\rm lop}^{R})}.
\label{equ:sig}
\end{equation}
Namely, the difference in the lopsided angle 
$\theta_{\rm lop}$ of a given system in the simulations and $\rm \langle\theta_{lop}^{R}\rangle$ in a uniform distribution is measured in units of the standard deviation of random distributions (which depends on $\nsate$). Note that $\rm \langle \theta_{lop}^{R} \rangle\approx 90^{\circ}$.

\begin{figure}[!ht]
\plotone{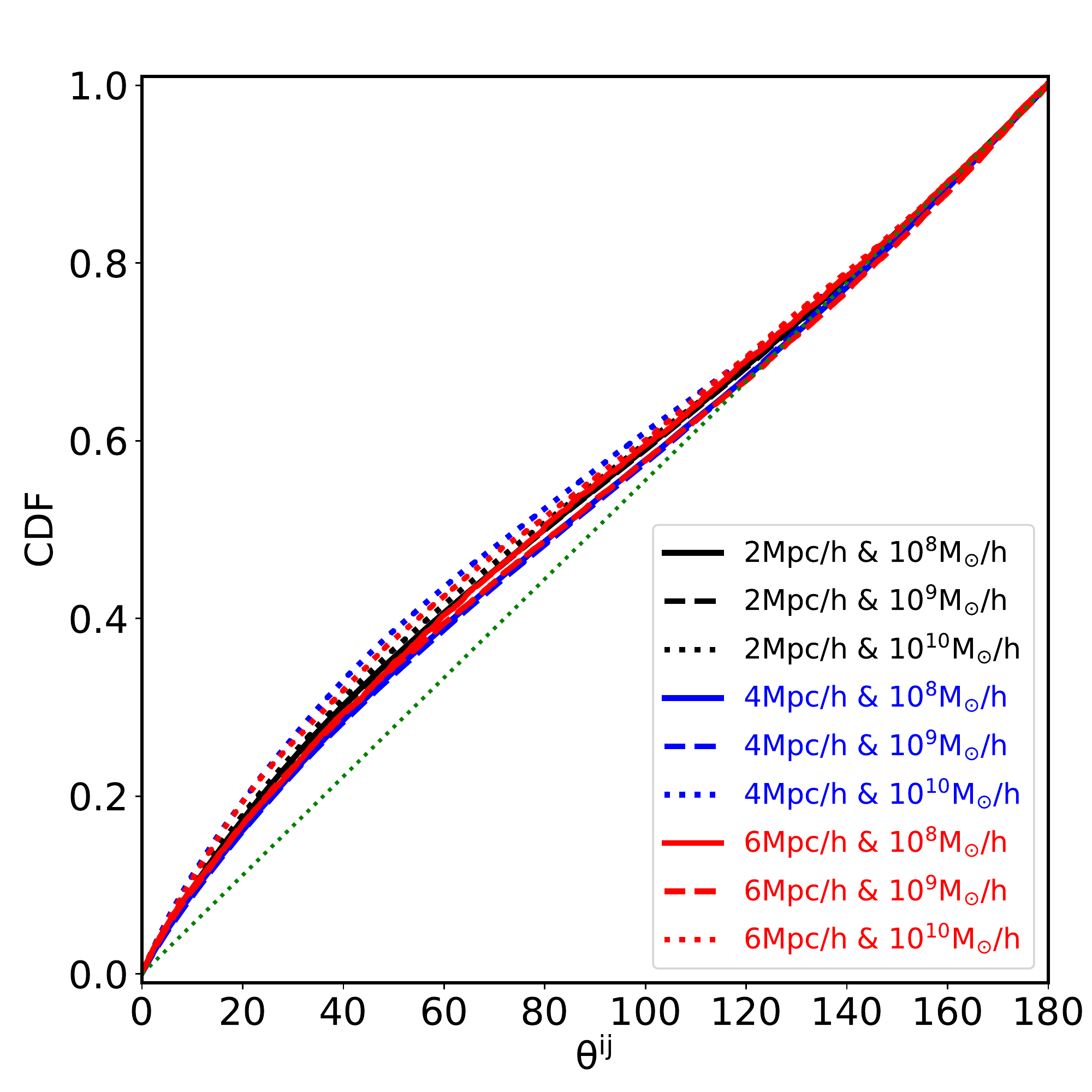}
\caption{{\bf The cumulative probability distribution of $\theta^{ij}$.} We examine the dependence of the lopsided signals on the various combinations between isolation distance ($2, 4$ and $6 \mpch$) and the lower mass limits of satellites ($10^8, 10^9$ and $10^{10}\msunh$). Note that for axis-symmetric satellite distribution, the expected distribution of $\theta^{ij}$ should follow the diagonal green line.}
\label{fig:lopsided_pdf}
\end{figure}

\section{Result}
\label{sec:result}

We begin the presentation of our results with two examples from the simulations, showing two satellite systems with strong and weak lopsidedness. Fig.~\ref{fig:lopsided_case} illustrates two cases with similar halo mass ($\sim10^{13.5}\msunh$) and satellite number ($\nsate$=14) but different angular distributions of satellites. In one case (shown in the left panel), satellite galaxies marked with black stars are entirely distributed on one side of the halo. The corresponding lopsided angle is $\theta_\mathrm{lop} =67.12^{\circ}$ and the lopsidedness fraction is $\rm f_{lop.}=1.0$ (i.e., 100\% of satellites reside on one side of the halo). We refer readers to App.~\ref{sec:flop_angle} for more details. Another extreme case is shown in the right panel of Fig.~\ref{fig:lopsided_case} in which the satellites are close to being isotropically distributed. Here, the lopsided angle $\rm \theta_\mathrm{lop}=92.69^{\circ}$ and the lopsidedness fraction is $\rm f_{lop.}=0.714$ (74.1\% of satellites reside on one side).

\subsection{The distribution of pairwise angles}
The simplest way to measure whether our sample described above has an axis-asymmetric satellite distribution is to examine the distribution of all angles formed between satellite pairs, namely the cumulative probability distribution of $\theta^{ij}$. This is shown in Fig.~\ref{fig:lopsided_pdf} for a variety of isolation criteria and mass cuts of satellites. We remind the reader that should the sample obey axis-symmetry, the curves in Fig.~\ref{fig:lopsided_pdf} would follow the diagonal (green dotted line). This is our first and main result: \textbf{\textit{Satellites in the Illustris-TNG simulation of the $\Lambda$CDM cosmology are distributed anisotropically and axis-asymmetrically around their (isolated) host dark matter halos.}} The median value of $\theta^{ij}\approx 80^{\circ}$\ is smaller than  $90^{\circ}$ expected for an uniform distribution. A quarter of these angles are less than $\sim$32$^{\circ}$, instead of less than $45^{\circ}$ expected for an uniform distribution. This is a clear demonstration that satellites are not symmetrically distributed around isolated hosts in $\Lambda$CDM cosmology, consistent with previous observational work \citepalias{2020ApJ...898L..15B}.

We note that our sample selection is based on two main parameters: the lower mass limit of satellites and the distance between hosts that defines isolated host galaxy. To assess how robust our findings are to these choices, various combinations of these two parameters have been examined:  three lower mass limits of $10^8$, $10^9$, and $10^{10}$ $\msunh$ have been used. Additionally, isolation distances of 2, 4, and 6 $\mpch$ were employed. As can be seen in Fig.~\ref{fig:lopsided_pdf} lopsidedness has a weak dependence on these parameters. Thus, for the rest of this paper, we use the fiducial values of $10^9$ $\msunh$ and 2 $\mpch$ for analyses.

\begin{figure*}[!ht]
\plotthree{Nsate_angle2d}{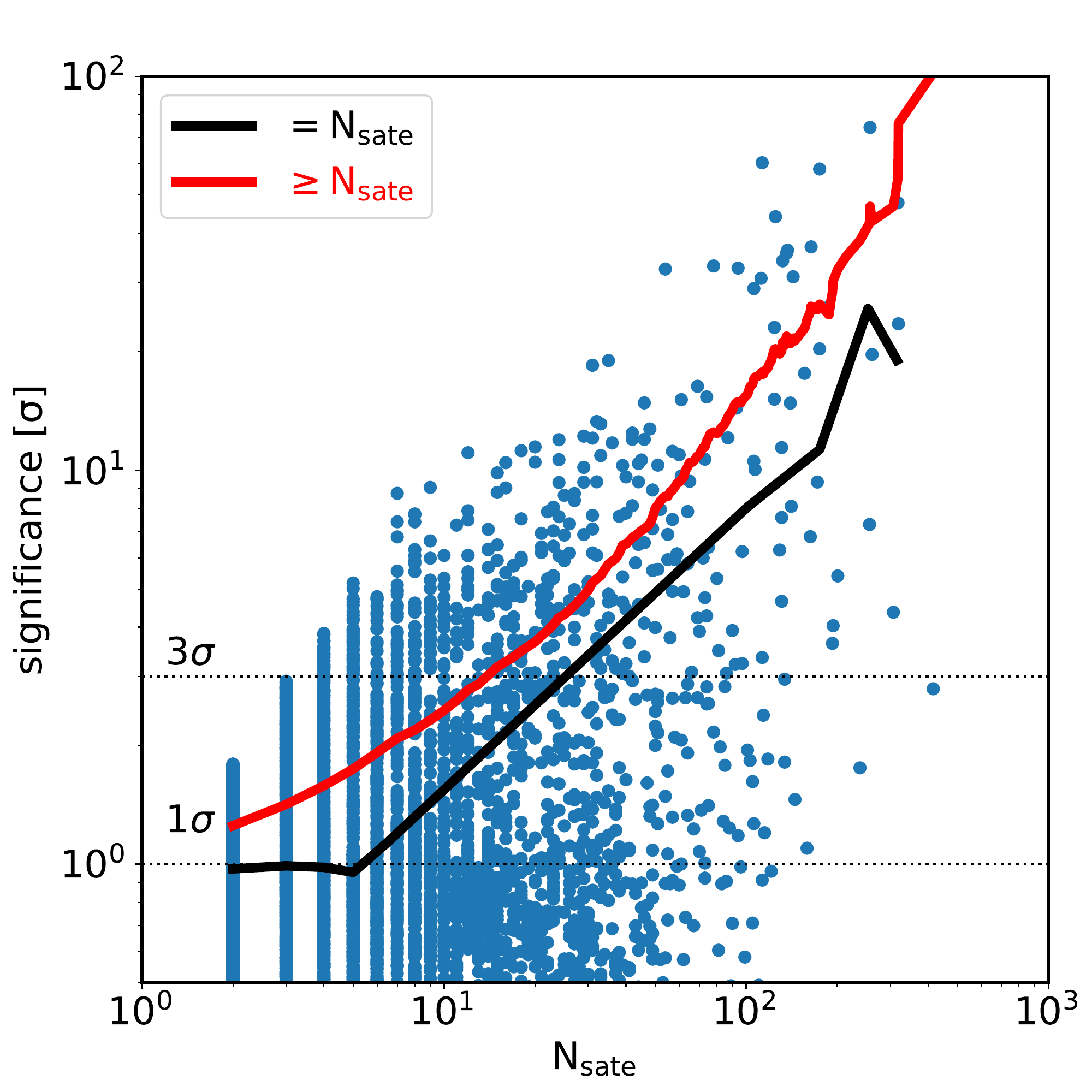}{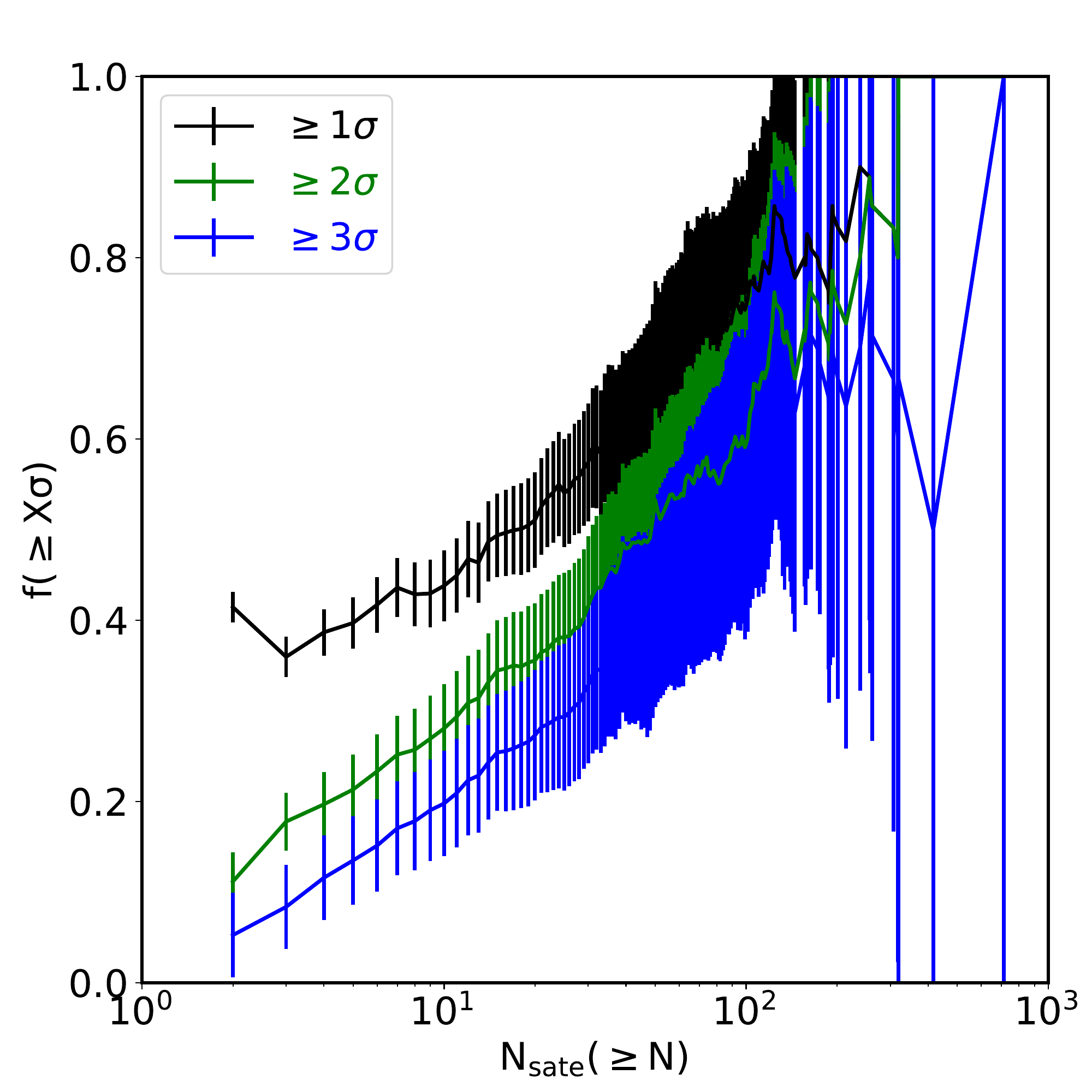}
\caption{{\bf Lopsided signal as a function of the number of satellites.} 
The {\bf left panel} displays the scatter distribution between the lopsided angle $\rm \theta_{lop}$ and $\nsate$. The red dashed line (=$90^{\circ}$) represents the expected angle for an isotropic satellite distribution, and the black dashed line represents the mean angle from the randomized trials. The color bands show the spread of $1\sigma$ (in grey), $2\sigma$ (in red) and $3\sigma$ (in yellow) deviation from 10,000 random uniform distributions.
{\bf Middle and Right panels} show the significance as a function of satellite number $\nsate$.  
In the middle panel, scatter points indicate each system's significance at the given number of satellites $\nsate$. The mean significance as a function of $\nsate$ ($\geq\nsate$) is shown in the solid black (red) line. Black dotted lines represent $1\sigma$ and $3\sigma$ confidence.  
The right panel shows the distribution of the number fraction of $N/N_{\rm tot}$, in which $N$ is the number of systems that contain at least $\nsate$ offset by at least a certain number of sigmas ($1\sigma$ in black, $2\sigma$ in green and $3\sigma$ in blue) from the randomized expectation. Note that the corresponding error bars are calculated by $1/\sqrt{N}$, in which $N$ is the number in each bin.}
\label{fig:lopsided_angle_sig}
\end{figure*}

\begin{figure*}[!ht]
\plotone{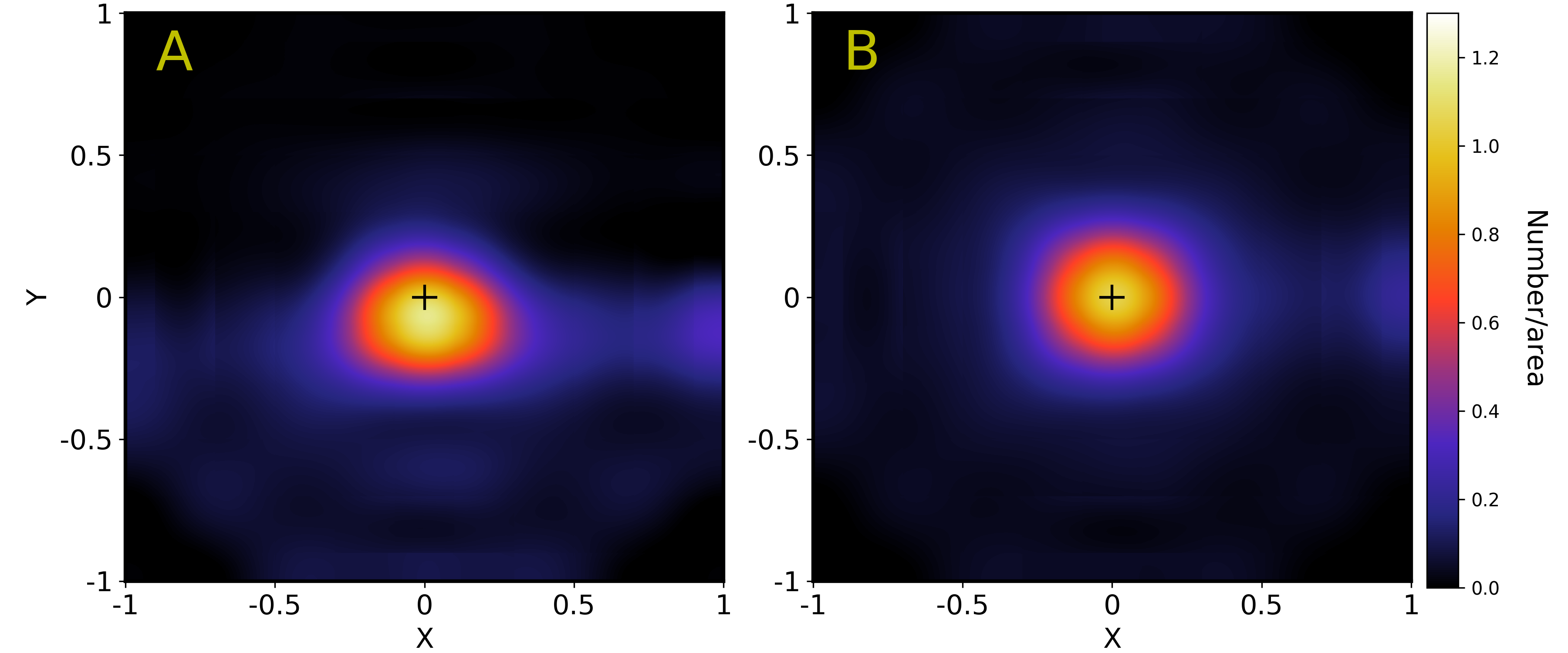}
\caption{{\bf The density distribution of stacked satellite galaxies in the $x$-$y$ plane.} The satellites' radial distances are scaled according to their distance to the central, normalized to the most distant satellite. Satellite systems are rotated such that the part containing more satellites faces the lower part of the Y-axis ({\bf Panel-A}). The random case is shown in  {\bf Panel-B}. The number of satellites in each arbitrary unit area is coded in the color bar. The black plus indicates the location of central galaxies.} 
\label{fig:lopsided_stacked}
\end{figure*}

\begin{figure*}[!ht]
\plotthree{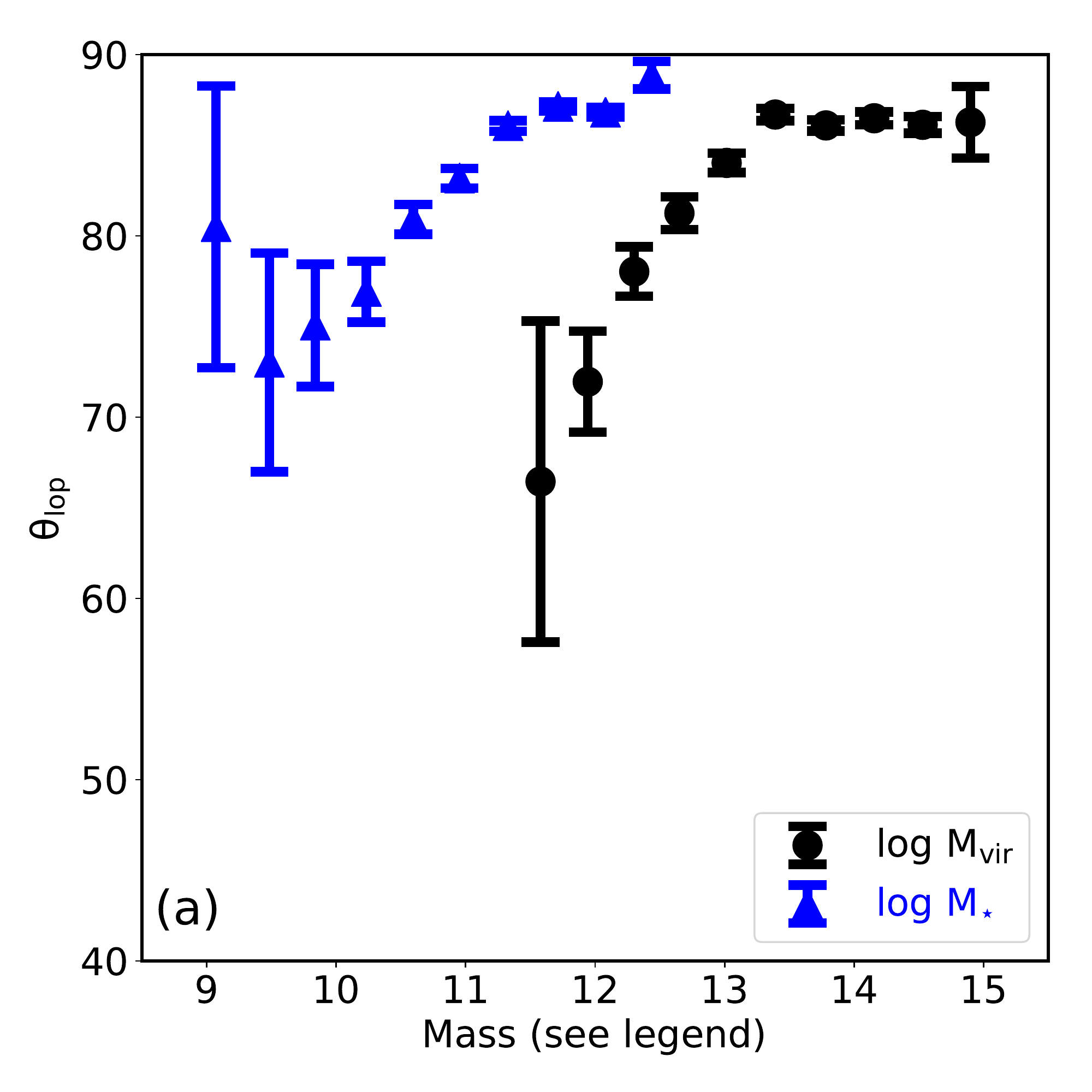}{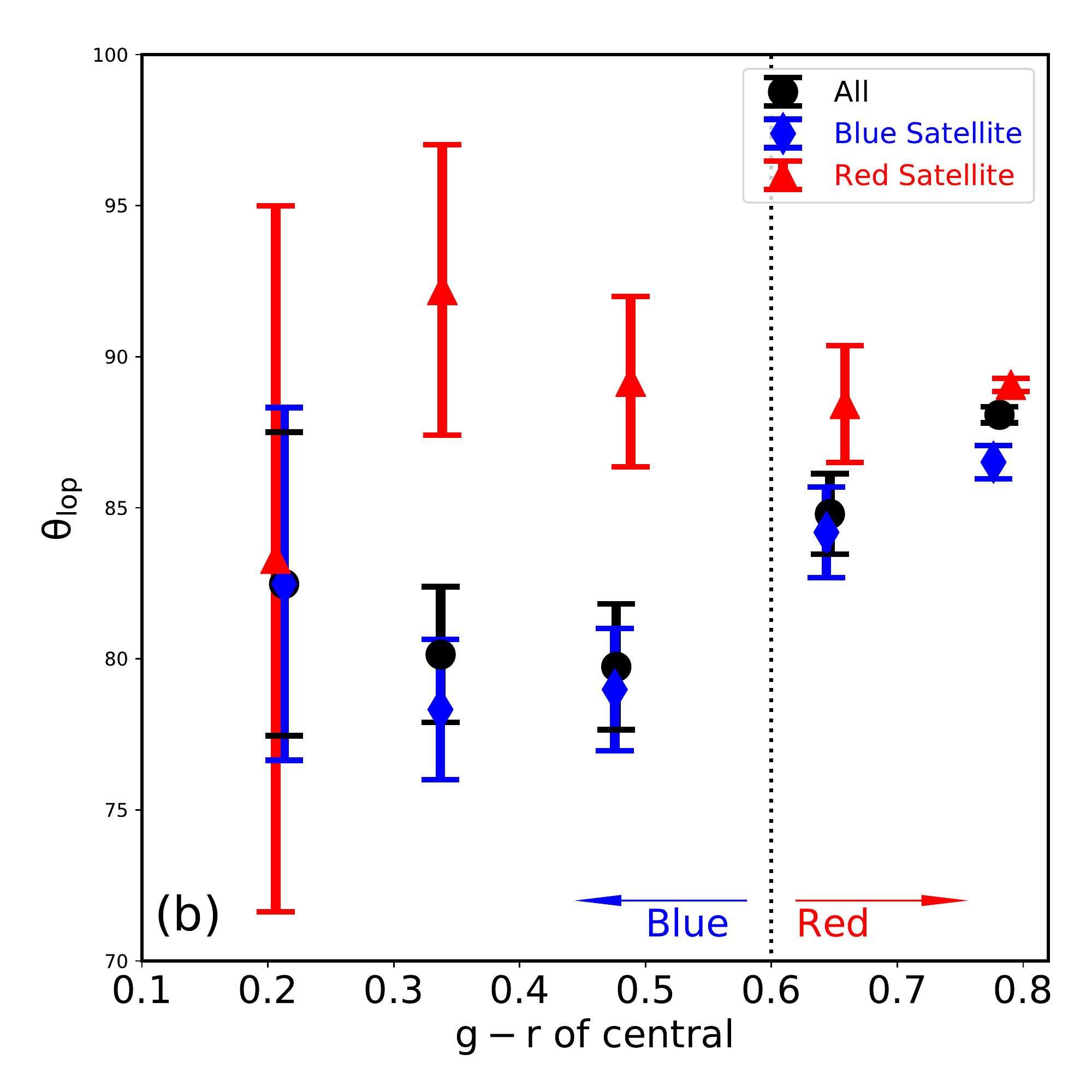}{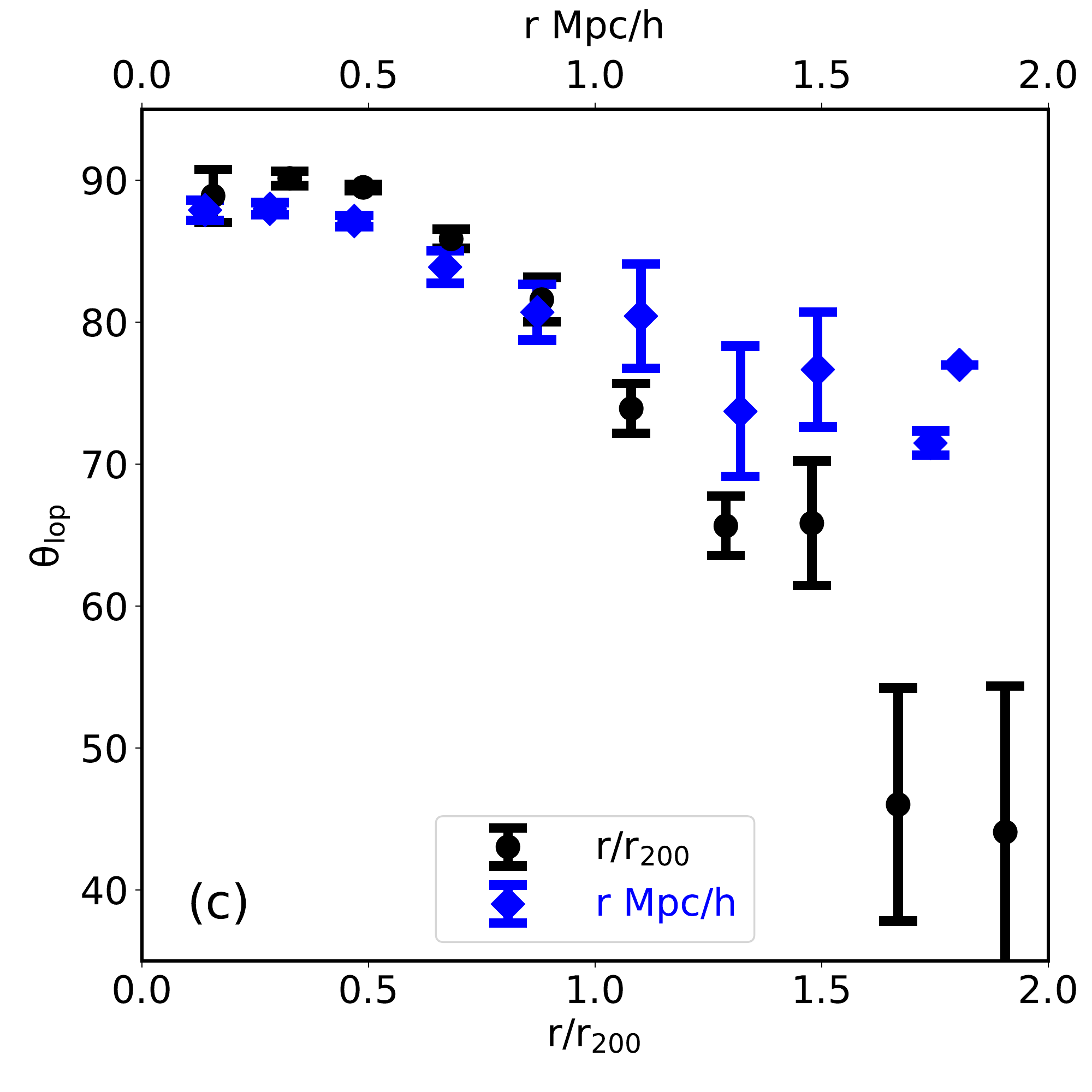}
\plotthree{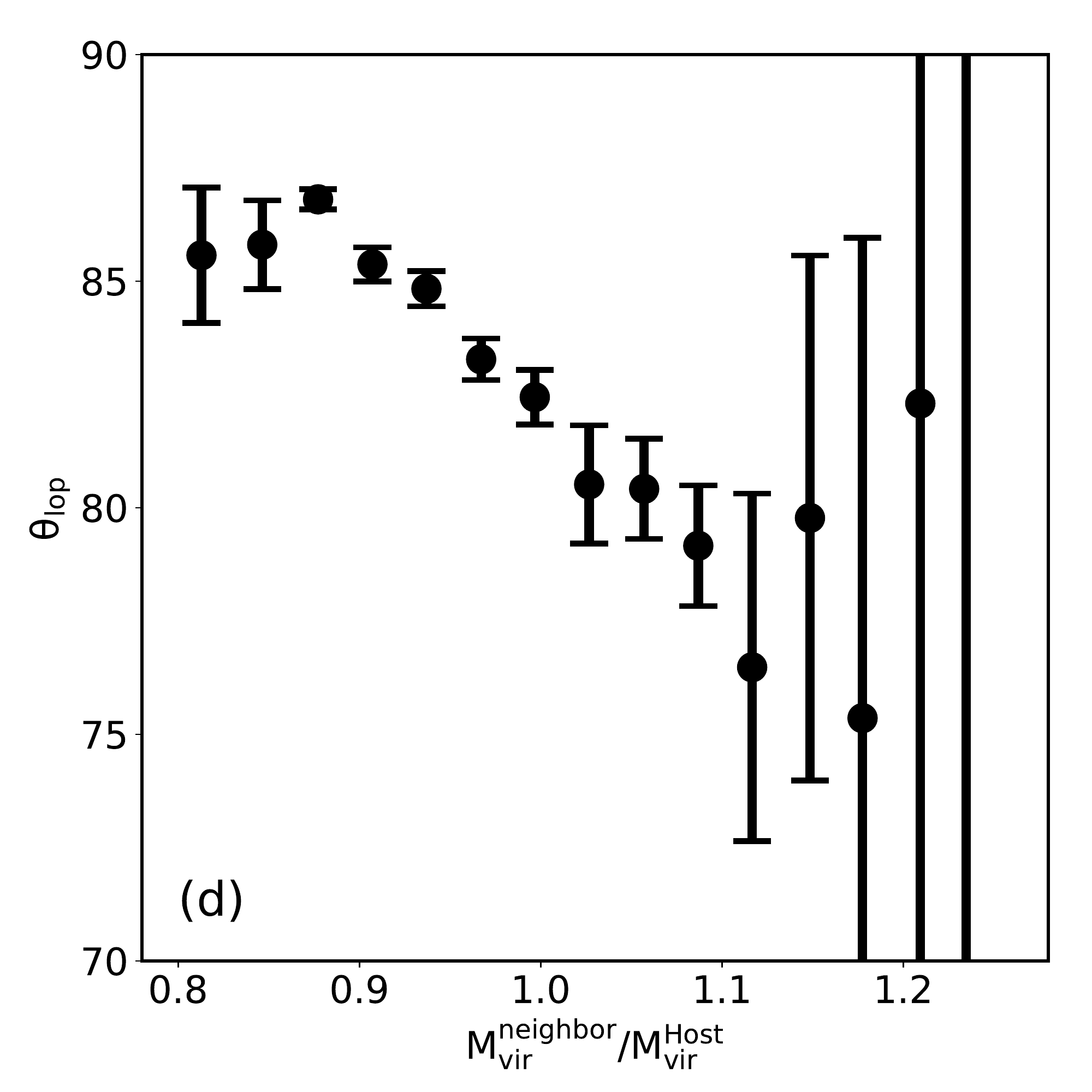}{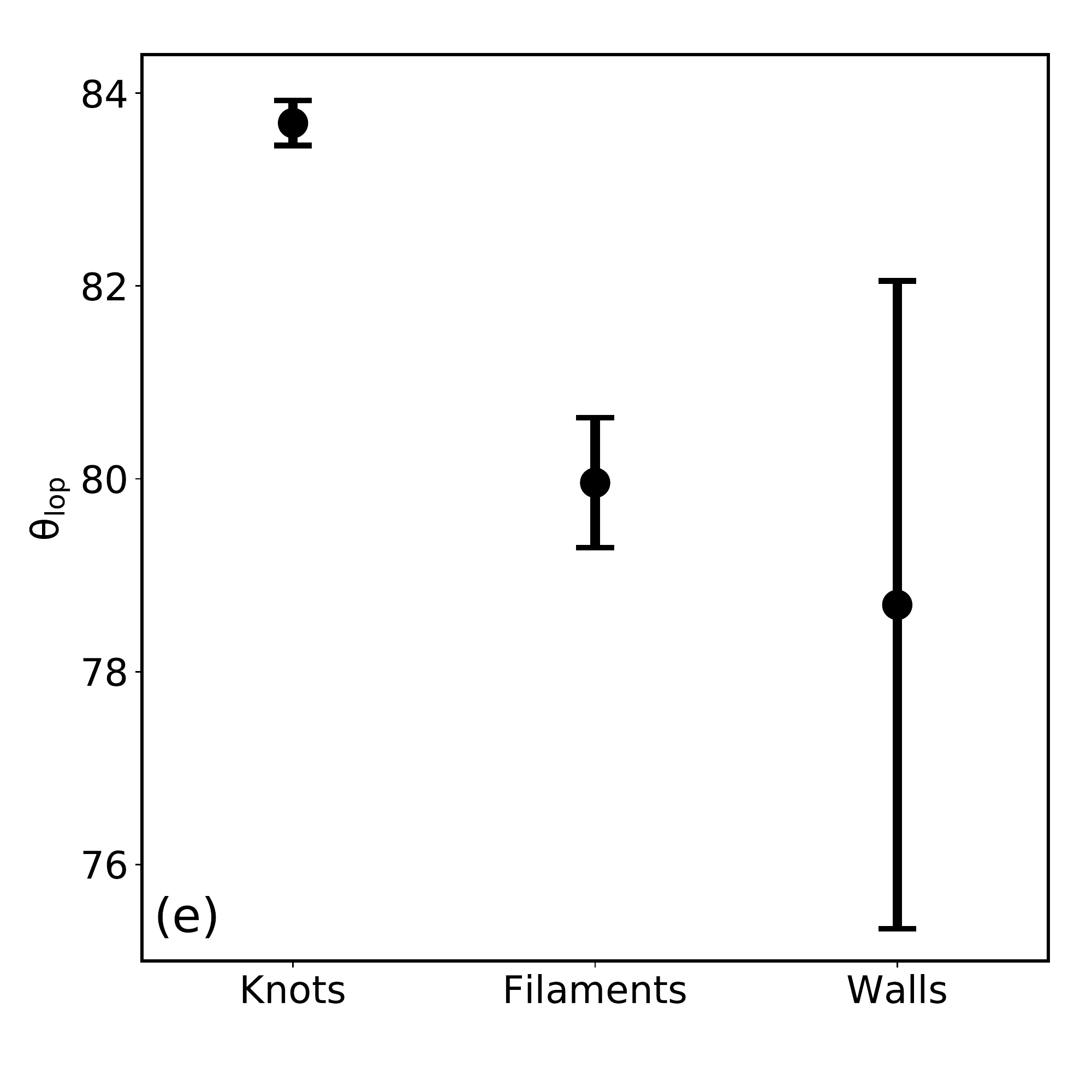}{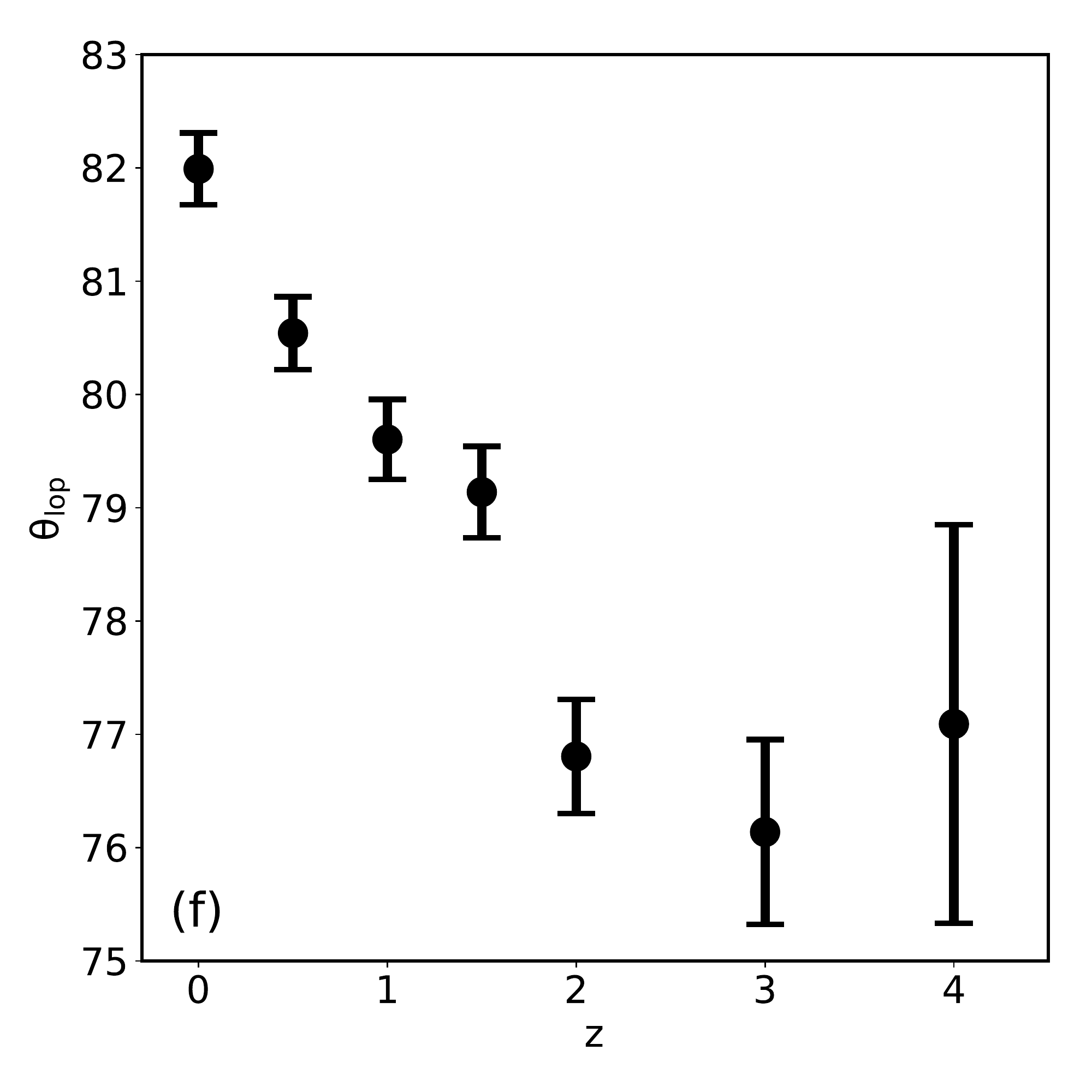}
\caption{{\bf The dependency of the lopsided angle on various properties of the centrals and satellites.}  
{\bf Upper panels a-c}: the dependency on the mass of the host, the color of the host and of the satellites, and the distance between the satellites and their central are examined.
Panel-a show the dependency of the lopsided angle on the host halo virial mass (in black-filled circles) and host galaxy stellar mass (in blue triangles).
Panel-b shows the dependency of the lopsided angle on the central g-r color. The `All' sample (black-filled circles) is divided into two subsamples according to the color of the satellites, shown in the red triangle and blue diamond, respectively. 
Panel-c represents the dependency on the distance between satellites and centrals. The black-filled circles are for distance normalized to the virial radius $r_{200}$ of the host, and blue diamonds for absolute distances.
{\bf Bottom panels d-f}: the dependency on the mass of neighbors, the cosmic web, and redshift are examined. 
Panel-d represents the dependency on the mass ratio between the closest neighbor and the host. 
Panel-e shows the large-scale environment dependence.
Panel-f examines the cosmic evolution of $\rm \theta_{lop}$. }
\label{fig:lopsided_properties}
\end{figure*}

\subsection{The distribution of lopsided angles}
As mentioned above, the significance of the lopsided signal is sensitive to the number of satellites in the host. To illustrate this we present the correlation between lopsided angle $\rm \theta_{lop}$, and $\nsate$ in the left panel of Fig.~\ref{fig:lopsided_angle_sig}. Each black dot denotes a single host system. The red dashed line (with $\rm \theta_{lop}$=$90^{\circ}$) represents the expected angle from an isotropic satellite distribution. The $1\sigma$, $2\sigma$, and $3\sigma$ spread standard deviations from 10,000 random realizations are shown in grey, red, and yellow, respectively. The mean value of lopsided angle $\rm \langle\theta^{R}_{lop}\rangle$ from the random realizations are shown as the black dashed line (this is limited by the number of random tests and will be close to the red dashed line at 90$^{\circ}$ if the number of the random tests is increased).  It can be clearly seen that there are a large number of systems for which  $\rm \theta_{lop}$ falls outside of the $3\sigma$ contours. The number of such significant systems clearly increases with increasing $\nsate$, indicating that the measured lopsided signal is clearly inconsistent with being random distribution for satellites.

The statistical significance of the measured lopsidedness is displayed in the middle and right panels of Fig.~\ref{fig:lopsided_angle_sig}. In the middle panel, we show the statistical significance as a function of $\nsate$. Again, each system is denoted by a single point. We note that although there are systems with many satellites consistent with axis-symmetry (i.e., $\sigma<1$), there is a trend wherein the larger $\nsate$, the more likely it is to be both lopsided and statistically significantly. The black line shows the mean significance of systems with the given number of $\nsate$, while the red line shows the mean significance of systems with at least the given number of $\nsate$. The significance increases with the increase of $\nsate$.  For the entire sample, the mean significance is slightly higher than $1\sigma$, and for systems with $\nsate\geq14$, the significance reaches up to $\sim3\sigma$. More interesting is that for $\nsate>100$, the significance of the lopsided signal is very high, up to $\sim10\sigma$. This may be a profound finding, given that Fig.~\ref{fig:sample} suggests that $\nsate$ correlates with halo mass: it implies that that the most massive halos are driving the lopsided signal, and thus (given that the most massive halos have the relatively smallest satellites) the lopsided distribution of satellite galaxies may be related to the mass ratio of satellites to hosts. More on this idea will be presented in the discussion section.

One may pose the question: what fraction of systems with at least $\nsate$ satellites are lopsided at the ``$n$'' $\sigma$ level? This is shown in the right panel of Fig.~\ref{fig:lopsided_angle_sig}, where the fraction of systems with lopsided distributions significant at the 1, 2 and 3 $\sigma$ level is shown, as a function of $\nsate$, by the black, green and blue lines, respectively. It is seen that, for instance, 35\% of all halos with at least two satellites are lopsided at the $1\sigma$ level, and around 50\% of halos with $\nsate\geq100$ are lopsided at a $3\sigma$ level.


The LSD of satellites in host galaxy can be visualized by stacking the entire sample and examining the projected (2D) number density of satellites. We place the halo center at the position of (x,y)=(0,0). The relative position of satellites to the center is scaled in units of the most distant satellite in each system, and the entire satellite system is rotated to ensure that the part containing more satellites faces the negative $y$ direction.  A significant excess is observed in the -y direction shown in the panel-A of Fig.~\ref{fig:lopsided_stacked}. However, in the cases of random samples shown in panel-B, no obvious lopsided signal was found, indicating that satellite galaxies are roughly evenly distributed around the central.

\subsection{Characteristics of LSDs}
We now move from quantifying the axis-asymmetrical nature of satellite distributions to examining their features and characteristics. In this section, we seek to understand the dependencies of the satellite distributions on the properties of the host and satellites themselves.  In Fig.~\ref{fig:lopsided_properties}, we present the median value of $\rm \theta_{lop}$ with dependence on the properties of the galaxy.  To assess the robustness of these estimates,  we compute the 68\% uncertainty intervals of the medians using bootstrapping. To do so, we randomly sample the systems within a given bin and calculate the median value of each re-sampled distribution. We repeat this process 1000 times, and adopt the dispersion of the corresponding medians as our uncertainty.

\subsubsection{Halo mass dependency}
The lopsided angle, $\rm \theta_{lop}$, of a given satellite system is shown as a function of stellar and halo mass of the host halo in blue triangles and black filled circles in Fig.~\ref{fig:lopsided_properties}a.  One distinctive feature of this plot sticks out: the lopsided signal is strongest (smallest $\rm \theta_{lop}$) at the low mass end: smaller halos have stronger LSDs than high-mass halos. However, the results is easy to be misinterpreted: the reason is due to the small number of satellites (relative larger error bars) in low mass halos. The lopsided signals of these systems with few satellites are more consistent with random and, as such, are less significant. As is visible, although the more massive halos -- by stellar or virial mass -- have $\rm \theta_{lop}$  that are closer to 90$^{\circ}$, these distributions are much more inconsistent with random (small error bars) than the low mass halos whose $\rm \theta_{lop}$ is further from  90$^{\circ}$.

\subsubsection{Color dependency}
Fig.~\ref{fig:lopsided_properties}b shows the dependency of the mean angle of LSD on the color of the host galaxy. The g-r color of the host galaxy is obtained from \cite{2018MNRAS.475..624N} who suggest that g-r$=$0.6 can be used to classify galaxies between red or blue. There are 1457 blue and 7267 red central galaxies in the sample of isolated hosts used here. 
As shown by the back-filled circles, a relatively weak dependence can be seen on g-r color: $\rm \theta_{lop}$ is independent (considering the large error range) with g-r for blue centrals, while a slight increase with g-r is seen for red centrals. 
The lopsided angles of all red centrals (with g-r$\geq$0.6) and all blue centrals (with g-r$<$0.6)  are 82.74$\pm$0.33 and 78.24$\pm$0.94, respectively. Blue centrals show a stronger lopsided strength than red central. This trend is in agreement with \citetalias{2020ApJ...898L..15B}. The lopsided signal of blue centrals is quantitatively in agreement with observations \citepalias[$79.65\pm0.91$ in][]{2020ApJ...898L..15B}, but red centrals in our study show stronger signal \citepalias[$89.05\pm0.22$ in][]{2020ApJ...898L..15B}. We also examined the impact of the satellites' color on the lopsided signal. It is found that at any given value of the x-axis (namely given the g-r color of central galaxies), the lopsided signal of red satellites (red triangles) is weaker than that of blue satellites (blue diamonds). A possible explanation of the color dependence is that the infall time of satellites around blue centrals is later. Thus, if a group of satellites are recently accreted, the system would have less time to be disrupted, and hence remain the initial lopsidedness.

\subsubsection{Radial dependency}
In Fig.~\ref{fig:lopsided_properties}c, the dependency of the lopsided angle on the satellite-central distance is presented. Satellite distance is scaled by the host halo's virial radius $r_{200}$. The black-filled circles show the dependence on the scaled distance and the blue diamonds show the dependence on the absolute physical distance. Note that halos outside of $r_{200}$ are still associated as satellites in several cases. This can happen because the Friends-of-Friends halo can be highly ellipsoidal (or ovoid), and as a result, objects that are bound to it may exist outside of $r_{200}$ (which is computed assuming sphericity). It can be seen in this figure that the axis-asymmetry is being driven primarily by the distant satellites. Although lopsidedness can be seen in the inner parts, there is a clear correlation with distance: the more distant a satellite is from the central, the stronger the lopsidedness.

Interestingly, there is little dependence of $\rm \theta_{lop}$ for satellites at distances greater than around 1 $\mpch$. This trend qualitatively agrees with the observational results by \citetalias{2020ApJ...898L..15B}, although they only examined satellites with projected distances less than 500 $\rm kpc$. The most pronounced difference is that \citetalias{2020ApJ...898L..15B} found that the LSDs signal is larger than $90^{\circ}$ within 100 $\rm kpc$. We speculate that this may be due to the following reasons. The main one is the difference in sample selection. We do not perform full mock-observations. Secondly, in the observation, some satellites may be located behind the central galaxy and thus cannot be observed, which may affect the lopsided signal. Thirdly, the difference may be caused by the difference between the galaxy formation model in the simulation or even the $\Lambda$CDM model and the real universe.

\subsubsection{Environment and redshift dependency}
Fig.~\ref{fig:lopsided_properties}d examines how the lopsided angle $\rm \theta_{lop}$ of a satellite system depends on the mass of the closest halo. Specifically the ratio $\rm M_{vir}^{neighbor}/M_{vir}^{Host}$ is used, in which $\rm M_{vir}^{neighbor}$ is the virial mass of the nearest neighbor. Note that by construction the nearest neighbor is at a distance $>2~\rm \mpch$. Fig.~\ref{fig:lopsided_properties}d indicates that the more massive the neighbor is, the more lopsided a system is. This finding is key in understanding what is driving the LSDs in halos. We speculate that the LSDs may be due to tidal effects from neighbors. Namely, the gravitational pull on satellite distributions from massive neighbors creates these lopsided distributions, skewing halos and turning spheres into ovoids \citep[inspired by][who explain the origin of LSDs in galaxy pairs]{2019MNRAS.488.3100G}.
This interpretation is entirely consistent with Fig.~\ref{fig:lopsided_properties}c, given that it is the most distant satellites, which are most susceptible to an external gravitational field, that are most lopsided.

Besides looking at the nearest neighbor's mass, the environment can also be classified according to the cosmic web. A review of such methods is presented in \cite{2018MNRAS.473.1195L}. In Fig.~\ref{fig:lopsided_properties}e the Hessian of the density field \citep{2020NewA...8001405W} is computed with a 2 $\mpch$ smoothing length. By counting the eigenvalues of the Hessian matrix, each halo can be ascribed to a particular cosmic web environment: knot, filament, wall, or void. Fig.~\ref{fig:lopsided_properties}e shows that $\rm \theta_{lop}$ decreases from knots to filaments and walls. This indicates that the strength of the lopsided signal, although with a large error bar in walls, increases as the environmental density decreases, similar to the trend shown in \cite{2018ApJ...859..115W} about the large-scale environment dependency of the satellite-central alignment. This can also be inferred from Fig.~\ref{fig:lopsided_properties}a since there exists a halo mass dependence on the cosmic web classification. Namely, the most massive halos tend to live in knots and the least massive ones in walls.

Finally, we investigate the redshift dependence of $\rm \theta_{lop}$ in Fig.~\ref{fig:lopsided_properties}f by performing our analysis at different redshifts from $z=0$ to $z=4$. It is seen that the lopsided angle increases with decreasing redshift, indicating that high LSDs are found at early Universe, and the LSD becomes weaker with time increase.  The dependence on redshift may be the result of multiple factors. Satellite systems become more virialized and relaxed (and hence more spherical), and the large-scale environments become denser as the density contrast increases.

\section{Summary and Discussion}\label{sec:sum_dis}
Lopsided distributions of satellite galaxies are not only found in galaxy pairs (\cite{2016ApJ...830..121L, 2017ApJ...850..132P, 2019MNRAS.488.3100G}),  but are also seen in isolated central galaxies in the data \citetalias{2020ApJ...898L..15B}. Motivated by the latter study based on the NASA-Sloan Atlas (NSA) catalog, we check if the lopsided distribution around isolated  is seen from the state-of-the-art hydrodynamical $\Lambda$CDM simulation.

In this paper, we investigated the (projected) lopsided distribution of satellite galaxies around isolated central galaxies in one hydrodynamic cosmological simulation. We found that the satellite distribution in isolated hosts is indeed lopsided, or asymmetric towards one side.  We summarize our conclusions in the below:

\begin{itemize}
    \item The axis-asymmetry or lopsidedness of satellite distribution can be measured in two ways. Either by looking at $\rm \theta^{ij}$, the distribution of all angles (Fig.~\ref{fig:lopsided_pdf}), centered on the host and subtended between satellites, or by looking at lopsided angle $\rm \theta_{lop}$ (the mean of $\rm \theta^{ij}$) for each halo. We found a statistically significant lopsided signal. 
    \item The statistical strength of the lopsided signal depends on the number of satellites $\nsate$ (see Fig.~\ref{fig:lopsided_angle_sig}). Small values of $\nsate$ (for example, systems with just 2 to 4 satellites) may exhibit strong lopsided distributions (i.e., both satellites are very close to each other and so $\rm \theta_{lop}$ is far from 90$^{\circ}$), but these are seldom statistically significant. For a system to have a statistically significant axis-asymmetric satellite distribution, it must have multiple satellites. Indeed, more satellites a system has, the greater the chance 
    that these are distributed with a significant degree of lopsidedness.
    \item The most massive halos are those that show the strongest lopsided signals (largest $\rm \theta_{lop}$ but most significant). 
    \item The satellites that are furthest from their hosts are the most lopsided (smallest $\rm \theta_{lop}$).
    \item Satellite systems of halos with the most massive neighbors exhibit the strongest lopsidedness. This result, together with the above result, indicate that it may be the tidal field that is responsible for asymmetrically compressing or expanding of the satellite systems.
    \item Satellite axis-asymmetry decreases as the universe evolves, likely due to the relaxation of halos and the more violent nature of the younger universe. 
\end{itemize}

Our results are generally consistent with those of \citetalias{2020ApJ...898L..15B}, especially in terms of color and satellite number dependence. The dependence of the large-scale environment and redshift indicate that the lopsided satellite distribution is the result of competition between the external environment (i.e., primordial infall) and evolution inside the halo.

The cold dark matter model predicts that  dark matter halo is a triaxial ellipsoid \citep{2002ApJ...574..538J}. For a relaxed (or close to relaxed) dark matter halo, the distribution of its satellite galaxies should coincide with the distribution of dark matter in the halo, such that the distribution of satellites is also in a triaxial ellipsoid.  However, our results might indicate that for systems with lopsided satellite distributions, the distribution of their hosts' dark matter is not triaxial but more ``ovoidal''. Suppose that the zeroth-order approximation is a spherical halo (i.e., one that follows an NFW profile) and the first-order description is an ellipsoidal halo (i.e., Jing \& Suto profile). In that case, our findings suggest that an ovoidal distribution is the second-order description of halo structure. It remains to be seen how such a correction affects the use of halo density profiles, for example, in the calculations of halo mass based on satellite dynamics.

It is interesting to ponder the physical origin of LSD in isolated central galaxies. Based on previous work about the impact of filaments on the satellites \citep{2019MNRAS.484.4325W,  2020ApJ...900..129W}, we speculate that this is closely related to the merger history and the large-scale environment. One possible explanation introduced by \cite{2018MNRAS.476.1796S} is that satellites of hosts at masses similar to the Milky Way are accreted in groups and mainly along the richest filament. In addition, \cite{2019MNRAS.488.3100G} found that the lopsided signal in central pairs originates from the highly anisotropic accretion of satellites, which may inform the case of isolated centrals. Moreover, it is generally accepted that satellites tend to be distributed around the major axis of the halo \citep[e.g.,][]{2006MNRAS.369.1293Y, 2018ApJ...859..115W}, however, these studies still imply axis-symmetry and thus can not explain the lopsided satellites distributions. No studies have shown that satellites are only distributed at one side of the halo major axis. But if this were the case, it would be another explanation for the lopsided distribution. Lastly, competition between the external environment (i.e., primordial infall) and the evolution of the halo \citep{2018ApJ...859..115W} may also be responsible for the lopsided signal seen in our work.

All the processes mentioned above are related to satellite accretion, since satellite accretion is close correlated with large scale structure \citep[e.g.,][]{2014MNRAS.443.1274L, 2015ApJ...807...37S, 2015ApJ...813....6K, 2017MNRAS.468L.123W, 2018MNRAS.473.1562W, 2020MNRAS.495..502M}, this implies that there is likely a  correlation between LSDs and how satellites are accreted. A more in-depth study that tracks the accretion of satellites should shed light on these issues.

\acknowledgments

We thank the anonymous referee for comments that substantially improved the manuscript.
PW thank Jiajun Zhang for useful discussion.
PW, NIL, XK, and QG acknowledge support from the joint Sino-German DFG research Project ``The Cosmic Web and its impact on galaxy formation and alignment'' (DFG-LI 2015/5-1, NSFC No. 1186131006). PW, MSP, and NL thank the DAAD for PPP grant 57512596 funded by the BMBF. MSP thanks the Klaus Tschira Stiftung gGmbH and German Scholars Organization e.V. for support via a Klaus Tschira Boost Fund. NIL acknowledges the financial support of Project IDEXLYON at the University of Lyon under the Investments for the Future Program (ANR-16-IDEX-0005). ET was supported by ETAg grant IUT40-2 and by EU through the ERDF CoE TK133. XK acknowledges financial support by the NSFC (No. 11825303, 11333008), the 973 program (No. 2015CB857003). QG acknowledges the financial support of the Shanghai Pujiang Program (No.19PJ1410700).


\appendix
\renewcommand\thefigure{\thesection\arabic{figure}}   
\setcounter{figure}{0}

\section{Lopsidedness fraction  versus lopsided angle}\label{sec:flop_angle}
\setcounter{figure}{0} 
For calculating $\rm f_{lop.}$,  one needs to draw a split line passing through the central galaxy to divide the projected plane into two parts and ensure one of them contains as many satellites as possible.  The $\rm f_{lop.}$ fraction is then defined as the ratio between the number of satellites on the side containing more satellites to the total number of satellites. Therefore, $\rm f_{lop.}$ ranges from 0.5 to 1.0.

\begin{figure}[!ht]
\plotone{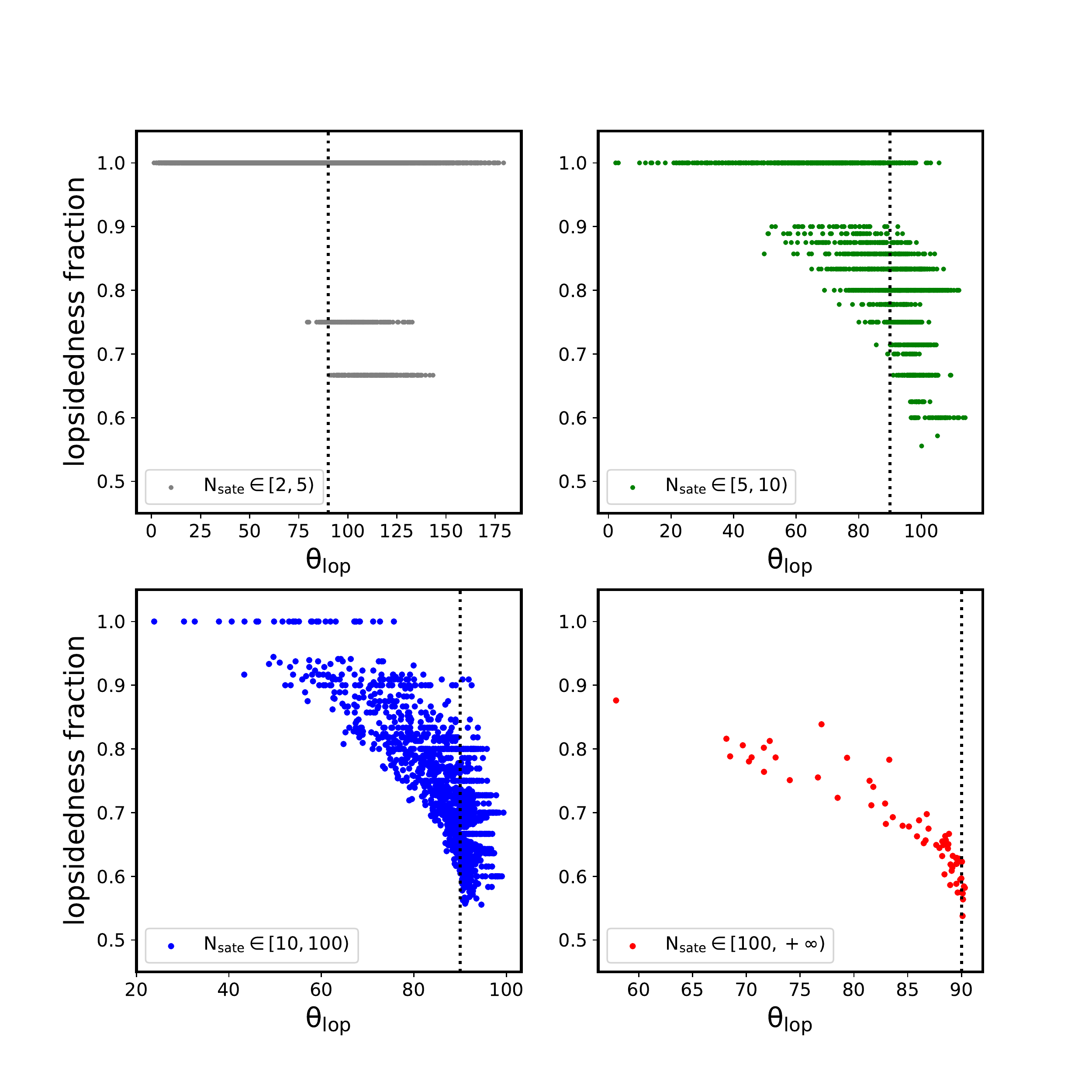}
\caption{The correlation between the lopsidedness fraction ($\rm f_{lop.}$) and lopsided angle ($\theta_{\rm lop}$) with considering the $\nsate$ effect. }
\label{fig:a1}
\end{figure}

The correlation between the lopsidedness fraction $\rm f_{lop.}$ and the lopsided angle $\theta_{\rm lop}$ is shown in Fig.~\ref{fig:a1}.  Two salient points can be gleaned here. First, generally, the lopsidedness fraction is positively correlated with the lopsided angle. However, for small $\nsate$, the correlation between $\rm f_{lop.}$ and $\theta_{\rm lop}$ is not clear (upper-left panel). For large $\nsate$, $\rm f_{lop.}$ increases (more lopsidedness) as $\theta_{\rm lop}$ decreases (more lopsidedness).  Second, even for a 100\% lopsided distributed satellite system ($\rm f_{lop.}$=1.0, i.e. all satellites are located on one side in the 2D plane), the $\theta_{\rm lop}$ may be greater than 90 degrees in some cases with small $\nsate$ (as shown in upper two panels and some points in the bottom-left panel). For very large $\nsate$, this situation does not happen. This indicates that the $\theta_{\rm lop}$ is not a perfect indicator to measure the lopsided distribution for an individual system with a small $\nsate$. For systems with large $\nsate$, $\theta_{\rm lop}$ can reliably indicate whether a given distribution is lopsided.

 \end{CJK*}

\begin{thebibliography}{10}
\bibitem[Agustsson \& Brainerd(2010)]{2010ApJ...709.1321A} Agustsson, I. \& Brainerd, T.~G.\ 2010, \apj, 709, 1321. doi:10.1088/0004-637X/709/2/1321
\bibitem[Ahn et al.(2014)]{2014ApJS..211...17A} Ahn, C.~P., Alexandroff, R., Allende Prieto, C., et al.\ 2014, \apjs, 211, 17. doi:10.1088/0067-0049/211/2/17
\bibitem[Azzaro et al.(2007)]{2007MNRAS.376L..43A} Azzaro, M., Patiri, S.~G., Prada, F., et al.\ 2007, \mnras, 376, L43. doi:10.1111/j.1745-3933.2007.00282.x
\bibitem[Bailin et al.(2008)]{2008MNRAS.390.1133B} Bailin, J., Power, C., Norberg, P., et al.\ 2008, \mnras, 390, 1133. doi:10.1111/j.1365-2966.2008.13828.x
\bibitem[Bennet et al.(2019)]{2019ApJ...885..153B} Bennet, P., Sand, D.~J., Crnojevi{\'c}, D., et al.\ 2019, \apj, 885, 153. doi:10.3847/1538-4357/ab46ab
\bibitem[Bennet et al.(2020)]{2020ApJ...893L...9B} Bennet, P., Sand, D.~J., Crnojevi{\'c}, D., et al.\ 2020, \apjl, 893, L9. doi:10.3847/2041-8213/ab80c5
\bibitem[Brainerd(2005)]{2005ApJ...628L.101B} Brainerd, T.~G.\ 2005, \apjl, 628, L101. doi:10.1086/432713
\bibitem[Brainerd \& Samuels(2020)]{2020ApJ...898L..15B} Brainerd, T.~G. \& Samuels, A.\ 2020, \apjl, 898, L15. doi:10.3847/2041-8213/aba194
\bibitem[Cautun et al.(2015)]{2015MNRAS.452.3838C} Cautun, M., Bose, S., Frenk, C.~S., et al.\ 2015, \mnras, 452, 3838. doi:10.1093/mnras/stv1557
\bibitem[Conn et al.(2013)]{2013ApJ...766..120C} Conn, A.~R., Lewis, G.~F., Ibata, R.~A., et al.\ 2013, \apj, 766, 120. doi:10.1088/0004-637X/766/2/120
\bibitem[Davis et al.(1985)]{1985ApJ...292..371D} Davis, M., Efstathiou, G., Frenk, C.~S., et al.\ 1985, \apj, 292, 371. doi:10.1086/163168
\bibitem[Dolag et al.(2009)]{2009MNRAS.399..497D} Dolag, K., Borgani, S., Murante, G., et al.\ 2009, \mnras, 399, 497. doi:10.1111/j.1365-2966.2009.15034.x
\bibitem[Georgiou et al.(2019)]{2019A&A...628A..31G} Georgiou, C., Chisari, N.~E., Fortuna, M.~C., et al.\ 2019, \aap, 628, A31. doi:10.1051/0004-6361/201935810
\bibitem[Gong et al.(2019)]{2019MNRAS.488.3100G} Gong, C.~C., Libeskind, N.~I., Tempel, E., et al.\ 2019, \mnras, 488, 3100. doi:10.1093/mnras/stz1917
\bibitem[Holmberg(1969)]{1969ArA.....5..305H} Holmberg, E.\ 1969, Arkiv for Astronomi, 5, 305
\bibitem[Ibata et al.(2013)]{2013Natur.493...62I} Ibata, R.~A., Lewis, G.~F., Conn, A.~R., et al.\ 2013, \nat, 493, 62. doi:10.1038/nature11717
\bibitem[Jing \& Suto(2002)]{2002ApJ...574..538J} Jing, Y.~P. \& Suto, Y.\ 2002, \apj, 574, 538. doi:10.1086/341065
\bibitem[Kang et al.(2007)]{2007MNRAS.378.1531K} Kang, X., van den Bosch, F.~C., Yang, X., et al.\ 2007, \mnras, 378, 1531. doi:10.1111/j.1365-2966.2007.11902.x
\bibitem[Kang \& Wang(2015)]{2015ApJ...813....6K} Kang, X. \& Wang, P.\ 2015, \apj, 813, 6. doi:10.1088/0004-637X/813/1/6
\bibitem[Libeskind et al.(2014)]{2014MNRAS.443.1274L} Libeskind, N.~I., Knebe, A., Hoffman, Y., et al.\ 2014, \mnras, 443, 1274. doi:10.1093/mnras/stu1216
\bibitem[Libeskind et al.(2016)]{2016ApJ...830..121L} Libeskind, N.~I., Guo, Q., Tempel, E., et al.\ 2016, \apj, 830, 121. doi:10.3847/0004-637X/830/2/121
\bibitem[Libeskind et al.(2018)]{2018MNRAS.473.1195L} Libeskind, N.~I., van de Weygaert, R., Cautun, M., et al.\ 2018, \mnras, 473, 1195. doi:10.1093/mnras/stx1976
\bibitem[Libeskind et al.(2005)]{2005MNRAS.363..146L} Libeskind, N.~I., Frenk, C.~S., Cole, S., et al.\ 2005, \mnras, 363, 146. doi:10.1111/j.1365-2966.2005.09425.x
\bibitem[Maji et al.(2017)]{2017ApJ...843...62M} Maji, M., Zhu, Q., Marinacci, F., et al.\ 2017, \apj, 843, 62. doi:10.3847/1538-4357/aa72f5
\bibitem[Marinacci et al.(2018)]{2018MNRAS.480.5113M} Marinacci, F., Vogelsberger, M., Pakmor, R., et al.\ 2018, \mnras, 480, 5113. doi:10.1093/mnras/sty2206
\bibitem[Morinaga \& Ishiyama(2020)]{2020MNRAS.495..502M} Morinaga, Y. \& Ishiyama, T.\ 2020, \mnras, 495, 502. doi:10.1093/mnras/staa1180
\bibitem[M{\"u}ller et al.(2018)]{2018Sci...359..534M} M{\"u}ller, O., Pawlowski, M.~S., Jerjen, H., et al.\ 2018, Science, 359, 534. doi:10.1126/science.aao1858
\bibitem[Nelson et al.(2018)]{2018MNRAS.475..624N} Nelson, D., Pillepich, A., Springel, V., et al.\ 2018, \mnras, 475, 624. doi:10.1093/mnras/stx3040
\bibitem[Pawlowski et al.(2012)]{2012MNRAS.423.1109P} Pawlowski, M.~S., Pflamm-Altenburg, J., \& Kroupa, P.\ 2012, \mnras, 423, 1109. doi:10.1111/j.1365-2966.2012.20937.x
\bibitem[Pawlowski et al.(2017)]{2017ApJ...850..132P} Pawlowski, M.~S., Ibata, R.~A., \& Bullock, J.~S.\ 2017, \apj, 850, 132. doi:10.3847/1538-4357/aa9435
\bibitem[Planck Collaboration et al.(2016)]{2016A&A...594A..13P} Planck Collaboration, Ade, P.~A.~R., Aghanim, N., et al.\ 2016, \aap, 594, A13. doi:10.1051/0004-6361/201525830
\bibitem[Sales \& Lambas(2004)]{2004MNRAS.348.1236S} Sales, L. \& Lambas, D.~G.\ 2004, \mnras, 348, 1236. doi:10.1111/j.1365-2966.2004.07443.x
\bibitem[Sastry(1968)]{1968PASP...80..252S} Sastry, G.~N.\ 1968, \pasp, 80, 252. doi:10.1086/128626
\bibitem[Shao et al.(2018)]{2018MNRAS.476.1796S} Shao, S., Cautun, M., Frenk, C.~S., et al.\ 2018, \mnras, 476, 1796. doi:10.1093/mnras/sty343
\bibitem[Shao et al.(2019)]{2019MNRAS.488.1166S} Shao, S., Cautun, M., \& Frenk, C.~S.\ 2019, \mnras, 488, 1166. doi:10.1093/mnras/stz1741
\bibitem[Shi et al.(2015)]{2015ApJ...807...37S} Shi, J., Wang, H., \& Mo, H.~J.\ 2015, \apj, 807, 37. doi:10.1088/0004-637X/807/1/37
\bibitem[Springel et al.(2001)]{2001MNRAS.328..726S} Springel, V., White, S.~D.~M., Tormen, G., et al.\ 2001, \mnras, 328, 726. doi:10.1046/j.1365-8711.2001.04912.x
\bibitem[Tang et al.(2020)]{2020ApJ...893...87T} Tang, L., Lin, W., \& Wang, Y.\ 2020, \apj, 893, 87. doi:10.3847/1538-4357/ab8292
\bibitem[Tully et al.(2015)]{2015ApJ...802L..25T} Tully, R.~B., Libeskind, N.~I., Karachentsev, I.~D., et al.\ 2015, \apjl, 802, L25. doi:10.1088/2041-8205/802/2/L25
\bibitem[van den Bosch et al.(2004)]{2004MNRAS.352.1302V} van den Bosch, F.~C., Norberg, P., Mo, H.~J., et al.\ 2004, \mnras, 352, 1302. doi:10.1111/j.1365-2966.2004.08021.x
\bibitem[Wan et al.(2020)]{2020MNRAS.492..456W} Wan, Z., Oliver, W.~H., Lewis, G.~F., et al.\ 2020, \mnras, 492, 456. doi:10.1093/mnras/stz3477
\bibitem[Wang et al.(2020)]{2020NewA...8001405W} Wang, P., Kang, X., Libeskind, N.~I., et al.\ 2020, \na, 80, 101405. doi:10.1016/j.newast.2020.101405
\bibitem[Wang et al.(2020)]{2020ApJ...900..129W} Wang, P., Libeskind, N.~I., Tempel, E., et al.\ 2020, \apj, 900, 129. doi:10.3847/1538-4357/aba6ea
\bibitem[Wang et al.(2019)]{2019MNRAS.484.4325W} Wang, P., Guo, Q., Libeskind, N.~I., et al.\ 2019, \mnras, 484, 4325. doi:10.1093/mnras/stz285
\bibitem[Wang \& Kang(2017)]{2017MNRAS.468L.123W} Wang, P. \& Kang, X.\ 2017, \mnras, 468, L123. doi:10.1093/mnrasl/slx038
\bibitem[Wang et al.(2018)]{2018ApJ...859..115W} Wang, P., Luo, Y., Kang, X., et al.\ 2018, \apj, 859, 115. doi:10.3847/1538-4357/aabe2b
\bibitem[Wang \& Kang(2018)]{2018MNRAS.473.1562W} Wang, P. \& Kang, X.\ 2018, \mnras, 473, 1562. doi:10.1093/mnras/stx2466
\bibitem[Yang et al.(2006)]{2006MNRAS.369.1293Y} Yang, X., van den Bosch, F.~C., Mo, H.~J., et al.\ 2006, \mnras, 369, 1293. doi:10.1111/j.1365-2966.2006.10373.x
\bibitem[York et al.(2000)]{2000AJ....120.1579Y} York, D.~G., Adelman, J., Anderson, J.~E., et al.\ 2000, \aj, 120, 1579. doi:10.1086/301513
\bibitem[Zhang \& Wang(2019)]{2019RAA....19..181Z} Zhang, M.-G. \& Wang, Y.\ 2019, Research in Astronomy and Astrophysics, 19, 181. doi:10.1088/1674-4527/19/12/181
\end{thebibliography}
\end{document}